%% file: 8593808.tex
\documentclass[journal]{IEEEtran}
\usepackage{array,multirow}
\usepackage{graphicx}
\usepackage{latexsym}
\usepackage{layout}
\usepackage{rotating}
\usepackage{bm}
\usepackage{euscript}
\usepackage{algorithmic}
\usepackage{amssymb, amsmath}
\usepackage[numbers, square, comma, sort&compress]{natbib}
\usepackage[ruled,lined]{algorithm2e}
\usepackage{epstopdf}
\usepackage[keeplastbox]{flushend}
\usepackage{gensymb}
\usepackage{subcaption}
\usepackage{siunitx}
\usepackage{url}
\usepackage{xcolor}
\usepackage{ragged2e}
\usepackage[english]{babel}
\usepackage{multicol}
\usepackage[official]{eurosym}
\usepackage{color}
\usepackage{acronym}
\usepackage{mathtools}
\usepackage[verbose]{wrapfig}
\usepackage{multirow}
\usepackage[normalem]{ulem}

\usepackage [autostyle, english = american]{csquotes}
\MakeOuterQuote{"}
\usepackage{caption} 
   \usepackage[belowskip=2pt, aboveskip=-13pt] {caption}
\DeclarePairedDelimiter{\ceil}{\lceil}{\rceil}

\newcommand{\eq}[1]{(\ref{#1})}

\acrodef{MEC}[MEC]{Multi-access Edge Computing}
\acrodef{EPC}[EPC]{Evolved Packet Core}
\acrodef{BS}[BS]{Base Station}
\acrodef{ML}[ML]{Machine Learning}
\acrodef{TIM}[TIM]{Telecom Italia Mobile}
\acrodef{EB}[EB]{Energy Buffer}
\acrodef{EH}[EH]{Energy Harvesting}
\acrodef{QoS}[QoS]{Quality of Service}
\acrodef{MN}[MN]{Mobile Network}
\acrodef{API}[API]{Application Programmable Interfaces}
\acrodef{App}[App]{Application}
\acrodef{ES}[ES]{Energy Savings}
\acrodef{NOES}[NOES]{NO Energy Saving}
\acrodef{VM}[VM]{Virtual Machine}
\acrodef{VUE}[VUE]{Virtual User Equipment} 
\acrodef{RNIS}[RNIS]{Radio Network Information Services} 
\acrodef{LOC}[LOC]{User Location Services} 
\acrodef{EM}[EM]{Energy Manager}
\acrodef{ENAAM}[ENAAM]{ENergy Aware and Adaptive Management}
\acrodef{LLC}[LLC]{Limited Lookahead Control}
\acrodef{RAN}[RAN]{Radio Access Network}
\acrodef{LSTM}[LSTM]{Long Short-Term Memory}
\acrodef{NFV}[NFV]{Network Function Virtualization}
\acrodef{VNF}[VNF]{Virtualized Network Function}
\acrodef{RNN}[RNN]{Recurrent Neural Network}
\acrodef{CPU}[CPU]{Central Processing Unit}
\acrodef{CH}[CH]{Cluster Head}
\acrodef{CM}[CM]{Cluster Member}

\begin{document}


\title{\ Online Supervisory Control and Resource Management for Energy Harvesting BS Sites Empowered with Computation Capabilities}

\author{\IEEEauthorblockN {Thembelihle Dlamini\IEEEauthorrefmark{1}\IEEEauthorrefmark{2}, \'Angel Fern\'andez Gamb\'in\IEEEauthorrefmark{1},   Daniele Munaretto\IEEEauthorrefmark{2}, Michele Rossi\IEEEauthorrefmark{1}}\\
	\IEEEauthorblockA {\IEEEauthorrefmark{1}Department of Information Engineering, University of Padova, Padova, Italy}\\
	\IEEEauthorblockA {\IEEEauthorrefmark{2}Athonet, Bolzano Vicentino, Vicenza, Italy}\\
	\{dlamini, afgambin, rossi\}@dei.unipd.it, daniele.munaretto@athonet.com \vspace{-0.4cm}
}

\maketitle
\thispagestyle{plain}
\pagestyle{plain}

\begin{abstract}
The convergence of communication and computing has lead to the emergence of {\it \ac{MEC}}, where computing resources (supported by \acp{VM}) are distributed at the edge of the \ac{MN}, i.e., in \acp{BS}, with the aim of ensuring reliable and \mbox{ultra-low} latency services. Moreover, \acp{BS} equipped with {\it\ac{EH}} systems can decrease the amount of energy drained from the power grid resulting into energetically \mbox{self-sufficient} \acp{MN}. The combination of these paradigms is considered here.  
Specifically,  we propose an online optimization algorithm, called \ac{ENAAM}, based on foresighted control policies exploiting (\mbox{short-term}) traffic load and harvested energy forecasts, where \acp{BS} and \acp{VM} are dynamically switched on/off towards energy savings and QoS provisioning. Our numerical results reveal that ENAAM achieves energy savings with respect to the case where no energy management is applied, ranging from $57\%$ and $69\%$. Moreover, the extension of ENAAM within a cluster of BSs provides a further gain ranging from $9\%$ to $16\%$ in energy savings with respect to the optimization performed in isolation for each \ac{BS}. 
\end{abstract}


\IEEEpeerreviewmaketitle

\section{Introduction}

The full potential of 5G radio access technology can be realized through the use of distributed intelligence, whereby content, control, and computation are moved closer to mobile users, hereby referred to as the {\it network edge}. This evolution has lead to the emergence of the \mbox{Multi-access} Edge Computing (MEC) paradigm, which allows network functions to be virtualized and then deployed at the network edge to guarantee the low latency required by some applications. In this paper, we consider a hybrid edge computing architecture where computing servers are \mbox{co-located} with each Base Station (BS), and a centralized controller (a point within range to a set of \acp{BS}) is utilized to manage them, deciding upon the allocation of their computing and transmission resources. This type of architecture is in line with recent trends~\cite{trends_huawei}.

The convergence of communication and computing (\ac{MEC}~\cite{etsimec}) within the mobile space poses new challenges related to energy consumption, as \acp{BS} are \mbox{densely-deployed} to maximize capacity and also empowered with computing capabilities to minimize latency. To cope with these challenges, previous studies have put forward \ac{BS} sleep modes~\cite{oh2011toward}\cite{superbowl}, as \acp{BS} are dimensioned for the expected maximum capacity, yet traffic varies during the day. In addition, energy savings within the virtualized computing platform are of great importance, as virtualization can also lead to energy overheads. Therefore, a clear understanding and a precise modeling of the server energy usage can provide a fundamental basis for server operational optimizations. The experimental results in~\cite{virttech}~\cite{eempirical} show that the locus of energy consumption for the \ac{VNF} components is the Virtual Machine (VM) instance where the \ac{VNF} is instantiated and executed. Thus, for a given expected traffic load, the energy consumption can be minimized by launching an optimal number of VMs, a technique referred to as \ac{VM} \mbox{\it soft-scaling}, together with \ac{BS} power saving methods, i.e., BS sleep modes.
 
Along these lines, we propose a \mbox{controller-based} network architecture for managing Energy Harvesting (EH) \acp{BS} empowered with computation capabilities where on/off switching strategies allow \acp{BS} and \acp{VM} to be dynamically switched on/off, depending on the traffic load and the harvested energy forecast, over a given lookahead prediction horizon. To solve the energy consumption minimization problem in a distributed manner, the {\it controller} partitions the \acp{BS} into clusters based on their location, then for each cluster, it minimizes a cost function capturing the individual communication site energy consumption and the users' \ac{QoS}. To manage the communication sites, the controller performs online supervisory control by forecasting the traffic load and the harvested energy using a \ac{LSTM} neural network~\cite{lstmlearn}, which is utilized within a \ac{LLC} policy (a predictive control approach~\cite{llcprediction}) to obtain the system control actions that yields the desired tradeoff between energy consumption and \ac{QoS}. 
This work is an extension of~\cite{online_pimrc}, where we consider energy savings within a single \mbox{off-grid} BS scenario (i.e., BS powered by either wind and solar energy sources) taking into account the need for MEC in remote/rural areas. In this paper, however, a dense environment is considered, similar to an urban or \mbox{semi-urban} scenario, where each BS is powered by hybrid energy supplies (solar and power grid) and empowered with computation capabilities. Moreover, the optimization problem is extended for multiple \acp{BS} where energy management procedures are executed within a \ac{BS} cluster in contrast with the single \ac{BS} case of~\cite{online_pimrc}.

The rest of the paper is structured as follows. The related work is discussed in Section~\ref{sec:related}, and the system model is presented in Section~\ref{sec:sys}. In Section~\ref{sec:case1}, we detail the optimization problem and the proposed \ac{LLC}-based online algorithm for a {\it single} communication site. The {\it multiple} BS communication site case is addressed in Section~\ref{sec:case2}. Our contribution is evaluated in Section~\ref{sec:results}, and lastly, concluding remarks are given in Section~\ref{sec:concl}.

\section{Related Work and Paper Contribution}
\label{sec:related}

Next, we first provide a literature review related to BS sleep modes techniques. Then, we review the mathematical tools that we use in this paper, followed by the literature review related to energy savings in virtualized computing platforms (i.e., works related to \mbox{soft-scaling}). Finally, we put forward our contributions and novelty of our work.\\ 

\noindent \textbf{\mbox{Sleep-mode} strategies in mobile networks:} cellular networks are dimensioned to support traffic peaks, i.e., the number of BSs deployed in a given area should be able to provide the required \ac{QoS} to the mobile subscribers during the highest load conditions. However, during \mbox{off-peak} periods the network may be underutilized, which leads to an inefficient use of network resources and to an excessive energy consumption. For these reasons, sleep modes have been proposed to dynamically \mbox{turn-off} some of the \acp{BS} when the traffic load is low. This has been extensively studied in the literature, here we highlight the main applied techniques that are related to this work.

Clustering algorithms have been proposed as a way of switching off \acp{BS} to reduce the energy consumption. In~\cite{Zhang2013energy}, centralized and distributed algorithms group \acp{BS} exhibiting similar traffic profiles over time. In~\cite{samarakoon2016dynamic}, a dynamic switching on/off mechanism locally groups \acp{BS} into clusters based on location and traffic load. The optimization problem is formulated as a \mbox{non-cooperative} game aiming at minimizing the BS energy consumption and the time required to serve their traffic load. Simulation results show energy costs and load reductions, while also providing insights of when and how the \mbox{cluster-based} coordination is beneficial.

Reducing the energy consumption involves some tradeoffs in the optimization problem. \ac{QoS} has been widely used as a tradeoff metric~\cite{Cai2013cross}~\cite{Zhu2014qos}. The Quality of Experience (QoE) is included in~\cite{Yuan2015qoe}, where a dynamic programming switching algorithm is put forward. Other parameters that have been considered are the coverage probability and the BS state stability parameter, i.e., the number of on/sleep state transitions. For instance, a set of \acp{BS} switching patterns engineered to provide full network coverage at all times, while avoiding channel outage, is presented in~\cite{Han2013energy}.  
According to the BS state stability concept, a \mbox{two-objective} optimization problem is formulated in~\cite{Liu2015base} and solved with two algorithms: (i) near optimal but not scalable, and (ii) with low complexity, based on particle swarm optimization.
The QoE is also affected by the UE position due to channel propagation phenomena. To this respect, in~\cite{Bousia2012ICC} the selection of the \acp{BS} to be switched off is taken so as to minimize the impact on the UEs' QoE, according to the distance from the handed off \acp{BS}.

To support sleep modes, neighboring cells must be capable of serving the traffic from the switched off cells. To achieve this, proper \textit{user association} strategies are required. A framework to characterize the performance (outage probability and spectral efficiency) of cellular systems with sleeping techniques and user association rules is proposed in~\cite{Tabassum2014downlink}. In that paper, the authors devise a user association scheme where a user selects its serving \ac{BS} considering the maximum expected channel access probability. This strategy is compared against the traditional maximum \mbox{SINR-based} user association approach and is found superior in terms of spectral efficiency when the traffic load is inhomogeneous. 
User association mechanisms that maximize energy efficiency in the presence of sleep modes are addressed in~\cite{Zhu2014energy}. There, a downlink HetNet scenario is considered, where the energy efficiency is defined as the ratio between the network throughput and the total energy consumption. Since this leads to a rather complex integer optimization problem, the authors propose a Quantum particle swarm optimization algorithm to obtain a suboptimal solution.

A marketing approach to foster the opportunistic utilization of the unexploited small cell (SC) BS capacity in dense heterogeneous networks (HetNets) is presented in~\cite{bousia2016multiobjective}. There, an offloading mechanism is introduced, where the operators lease the capacity of a SC network owned by a third party in order to switch off their BSs (Macro BSs) and maximize their energy efficiency, when the traffic demand is low. The allocation of the SC resources among a set of competing operators is mathematically formulated as an auction problem.

A comprehensive power management model employing a BS switching on/off mechanism, within a BS system powered by green energy, is presented in~\cite{bs_energy_cost}. The model considers weather conditions, user mobility, different green energy harvesting rates, energy storage with \mbox{self-discharge} effect, and switching on/off frequency. The authors propose two algorithms: the first decides which \acp{BS} are to be active based on the minimum energy cost, i.e., the energy price per time period, while the second one determines the active \acp{BS} by first prioritizing the minimum power consumption of the system, and then the energy cost.
The relationship between installing a solar harvesting system to power a \ac{BS} and the energy management under varying demand is investigated in~\cite{planning_solar}. The authors present a solar installation planning model by explicitly modelling solar panels, batteries, inverters and charge controllers, as well as the cellular network demand and energy management. They found that the solar installation and the energy management of the base stations are so coupled that even the order in which these technologies are introduced can have a major impact on the network cost and performance.

The survey paper~\cite{angel_sustainable}, presents a taxonomy of existing energy sustainable paradigms and methods to address energy savings in network elements (i.e., BSs) equipped with EH capabilities. Here, the authors discuss the shortcomings of previous studies related to efficient energy management procedures, the lack of relevant discussion related to the integration of EH into future networks, and lastly, energy \mbox{self-sustainability} in future networks. The current work is a technical contribution where we address some of the shortcomings that were identified in~\cite{angel_sustainable}, also proposing the use of Machine Learning (ML) tools for pattern forecasting and adaptive control schemes for decision making. In addition, this work is in line with the research topics which can be found in our review paper~\cite{energymanagershow}.

The majority of the works on BS switching off mechanism considered clusters of BSs from a {\it single} mobile operator perspective, where some functions of the BS can be switched off and then the remaining active BSs handle the upcoming traffic. A new approach is presented in~\citep{antonopoulos2015energy} which exploits the coexistence of multiple BSs from different mobile operators in the same area. An \mbox{intra-cell} \mbox{roaming-based} \mbox{infrastructure-sharing} strategy is proposed, followed by a distributed \mbox{game-theoretic} \mbox{switching-off} scheme that takes into account the conflicts and interaction among the different operators. Moreover, in~\cite{oikonomakou2017evaluating}, the authors investigate the energy and cost efficiency of multiple HetNets (i.e., each HetNet is composed of eNodeBs (eNBs) and SC BSs from one operator) that share their infrastructure and also are able to switch off part of it. Here, a form of \mbox{roaming-based} sharing is also adopted, whereby the operator can roam its traffic to a rival operator during a predefined period of time and area. An energy efficient optimization problem is formulated and solved using a cooperative greedy heuristic algorithm. Regarding the cost efficiency, the cooperation and cost sharing decisions among the operators are modeled using a Shapley Value based bankruptcy game.\\

\noindent \textbf{Pattern forecasting along with foresighted optimization:} \mbox{control-theoretic} and \ac{ML} methods for resource management have been successfully applied to various problems, e.g., task scheduling, bandwidth allocation, network management policies, etc. In the paradigm of supervisory control for managing Mobile Networks (MNs), online forecasting using \ac{ML} techniques and the \ac{LLC} method can yield the desired system behavior when taking into account the environmental expectations, i.e., traffic load and energy to be harvested. Next, we briefly review the mathematical tools that we use in this paper, namely the \ac{LLC} method and \ac{LSTM} neural network~\cite{lstmlearn}. 

\mbox{Control-theorectic} algorithms and the \ac{LLC} method have been used to obtain control actions that optimize the system behavior, by employing a forecasting mathematical model, over a limited \mbox{look-ahead} prediction horizon. \ac{LLC} is conceptually similar to Model Predictive Control (MPC)~\cite{mpc_book}.
In~\cite{chung1992limited}, an online supervisory control scheme based on \ac{LLC} policies is proposed. Here, after the occurrence of an event, the next control action is determined by estimating the system behavior a few steps into the future, using the currently available information as inputs. The control action exploration is performed using a search tree assuming that the controller knows all future possible states of the process over the prediction horizon. Moreover, in~\cite{llcprediction}, an online control framework for resource management in switching hybrid systems is proposed, where the system's control inputs are finite. The relevant parameters of the operating environment, e.g., workload arrival, are estimated and then used by the system to forecast future behavior over a \mbox{look-ahead} horizon. From this, the controller optimizes the predicted system behavior following the specified \ac{QoS} through the selection of the system controls. 

To model \mbox{time-series} datasets, the \ac{LSTM} network is used as it is able to handle the \mbox{long-term} dependencies due to its inherent capability of storing past information and then recalling it. 
In~\cite{ergen2017online}, a distributed \ac{LSTM} online method based on the particle filtering algorithm is presented with an aim of investigating the performance of online training of \ac{LSTM} architectures in a distributed network of nodes. An \ac{LSTM} based model for variable length data regression is proposed, and then put into a nonlinear \mbox{state-space} form to train the model in an online fashion. Then, financial and real life datasets are used for performance evaluation, and it is observed that the distributed online approach yields the same results that are obtained in the centralized case, when considering the mean square errors as the performance measure. 
Moreover, an \ac{LSTM} forecasting method is utilized in~\cite{online_pimrc} within an \mbox{LLC-based} algorithm to obtain the system control actions yielding the desired tradeoff between energy consumption and \ac{QoS}, for a remote site powered by only green energy.\\

\noindent \textbf{Energy savings in virtualized platforms through \mbox{soft-scaling}:} with the advent of virtualization, it is expected that the \ac{NFV} framework can exploit the benefits of virtualization technologies to significantly reduce the energy consumption of large scale network infrastructures. In virtualized computing environments, the locus of energy consumption for components is due to the \acp{VM} running in the server(s). Thus, energy saving studies within the virtualized computing environment have involved the scaling down of the number of computing nodes/servers (autoscaling~\cite{xu2016online}), \ac{VM} migration~\cite{Beloglazov} (movement of a \ac{VM} from one host to another) and soft resource scaling~\cite{soft_virtual} (shortening of the access time to physical resources), all hereby referred to as \ac{VM} \mbox{\it soft-scaling}, i.e., the reduction of computing resources per time instance. 

Algorithms for the dynamic on/off switching of servers have been proposed as a way of minimizing energy consumption in computing platforms. In~\cite{xu2016online}, at the beginning of each time slot computing resources are provisioned depending on the expected server workloads via a reinforcement \mbox{learning-based} resource management algorithm, which learns \mbox{on-the-fly} the optimal policy for dynamic workload offloading and the autoscaling of servers. 
Then in~\cite{online_pimrc}, computing resources (\acp{VM}) are provisioned based on a \ac{LLC} policy after forecasting the future workloads and harvested energy.
In~\cite{Beloglazov}, the \ac{CPU} utilization thresholds are used to identify \mbox{over-utilized} servers. Hence, migration policies, enabled by the live \ac{VM} migration method~\cite{livemig}, are applied for moving the \acp{VM} between physical nodes (servers). The \acp{VM} are only moved to hosts that will accept them without incurring high energy cost, i.e., without any increase in the \ac{CPU} utilization. Subsequently, the idle servers are switched off. 

Power management is also of interest in virtualized computing platforms, i.e., data centers using virtualization technologies. In~\cite{soft_virtual}, a power management approach called {\it VirtualPower} is presented. The algorithm exploits hardware power scaling, i.e., the dynamic power management strategies using Dynamic Voltage and Frequency Scaling (DVFS)~\cite{server_manager}\cite{coca_dvfs}, and \mbox{software-based} methods, i.e., scaling the allocation of physical resources to \acp{VM} using the hypervisor scheduler, for controlling the power consumption of underlying platforms. Due to the low power management benefits obtained from hardware scaling, a {\it soft resource scaling} mechanism is proposed whereby the scheduler shortens the maximum resource usage time for each \ac{VM}, i.e., the time slice allocated for using the underlying physical resources. \\

\begin{figure}[t]
	\centering
	\includegraphics[width =\columnwidth]{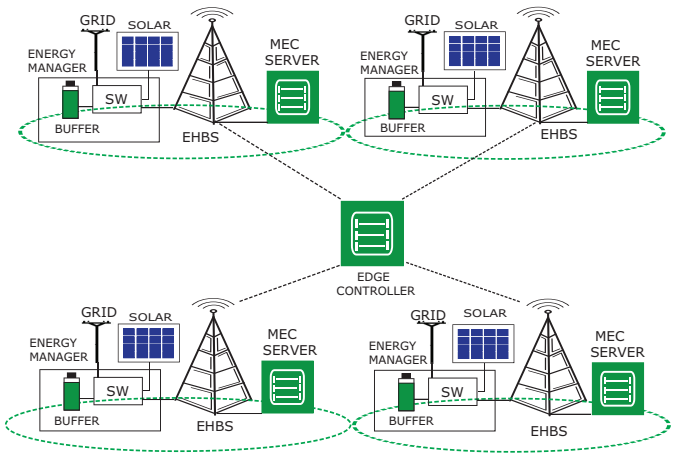}
	\caption{Edge network topology. The electromechanical switch (SW) selects the appropriate source of energy.}
	\label{fig:mecscenario}
\end{figure} 

\noindent \textbf{Novelty of this work:} here, we consider the aforementioned scenario, where each \ac{BS} is equipped with \ac{EH} hardware (a solar panel for \ac{EH} and an \ac{EB} for energy storage) and a \ac{MEC} server \mbox{co-located} with the BS for computation purposes, under the management enabled by the {\it controller}.

Motivated by the potential capabilities of \ac{EH}, \ac{MEC} and the presence of the controller, 
\begin{itemize}
\item[1)] we introduce the use of virtualization with the aim of investigating how \acp{VM} can be \mbox{soft-scaled} based on the forecasted server workloads, as \acp{VM} are the source of energy consumption in computing environments.
\item[2)] We put forward the edge \mbox{controller-based} architecture for small cell \acp{BS} management, as one of the future trends for small cells~\cite{trends_huawei} in 5G MNs. 
\item[3)] We reconsider the BS sleeping control mechanism under the new MEC paradigm, which has not been sufficiently covered in the literature. In addition, we use a clustering method for enabling energy savings within the MN. 
\item[4)] We estimate the \mbox{short-term} future traffic load and harvested energy in \acp{BS}, by using \ac{LSTM} neural network~\cite{forecasting}. 
\item[5)] We develop an online supervisory control algorithm for the radio access (edge) network management based on a predictive method, specifically the \ac{LLC} method, along with clustering and energy management procedures. The main goal is to enable \ac{ES} strategies within the access network, \ac{BS} sleep modes and \ac{VM} \mbox{soft-scaling}, following the energy efficiency requirements of a virtualized infrastructure from~\cite{eerequirements}. The proposed management algorithm is called ENergy Aware and Adaptive Management (ENAAM) and is hosted in the edge controller. The \ac{ENAAM} algorithm considers the future BS traffic load, onsite green energy in the \ac{EB} and then provisions access network resources, per communication site, based on the learned information, i.e., energy saving decisions are made in a \mbox{forward-looking} fashion.
\end{itemize}

The proposed optimization strategy leads to a considerable reduction in the energy consumed by the edge computing and communication facilities, promoting \mbox{self-sustainability} within the mobile network through the use of green energy. This is achieved under the controller guidance, which makes use of forecasting, clustering, control theory and heuristics.

\section{System Model}
\label{sec:sys}

\begin{figure}[t]
	\centering
	\resizebox{\columnwidth}{!}{\input{./traffic_profiles.tex}}
	\caption{Example traces for normalized BS traffic loads. The data from~\cite{bigdata2015tim} has been split into four representative clusters.}
	\label{fig:trace_load}
\end{figure}
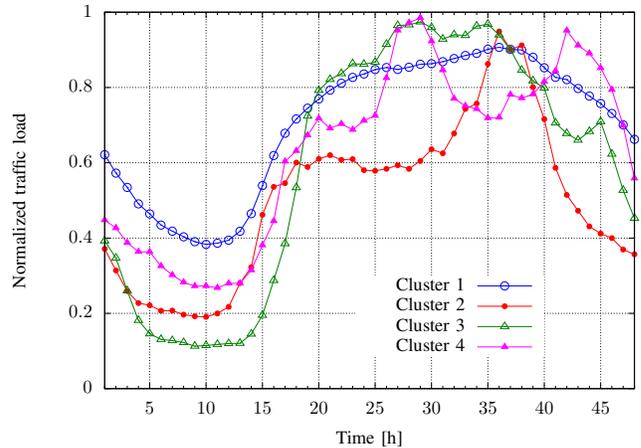 

As a major deployment of MEC and in line with current trends for future mobile networks as suggested by prominent network operators (e.g., Huawei Technologies~\cite{trends_huawei}), the considered scenario is illustrated in Fig.~\ref{fig:mecscenario}. It consists of a \mbox{densely-deployed} \ac{MN} featuring  $N$ \acp{BS} and \mbox{co-located} \mbox{cache-enabled} MEC servers. Each MEC server hosts $M$ \acp{VM}. Each communication site, i.e., the \ac{BS} and the \mbox{co-located} MEC server, is empowered with \ac{EH} capabilities through a solar panel and an \ac{EB} that enables energy storage. Energy supply from the power grid is also available. Moreover, the \ac{EM} is an entity responsible for selecting the appropriate energy source and for monitoring the energy level of the \ac{EB}. All \acp{BS} communicate with a centralized entity called the {\it edge controller}, which is responsible for managing the access network apparatuses. The energy level information is reported periodically to the edge controller through the pull file transfer mode procedure (e.g., File Transfer Protocol~\cite{filetransfer}). Moreover, we consider a \mbox{discrete-time} model, whereby time is discretized as $t = 1,2,\dots$, and each time slot $t$ has a fixed duration $\tau$. The list of symbols that are used in the paper is reported in Table~\ref{tab:variables}. 

\begin{table*}[t]
\caption{Notation: list of symbols used in the analysis.}
\center
\begin{tabular} {|l|l|}
\hline 
{\bf Symbol} & {\bf Description} \\ 
\hline
 & {\bf \qquad\qquad\qquad\qquad Input Parameters}\\
\hline
$N$ & number of \acp{BS}, indexed by $n$\\
$M$ & maximum number of \acp{VM} hosted by each MEC server\\
$\tau$ & time slot duration \\
$L_{n}(t)$ & BS $n$ traffic load profile in time slot $t$, $n$ is the BS index\\
$\Gamma_{n}(t)$ & workload handled by the MEC server at BS $n$ in time slot $t$\\
$\Gamma^\prime_{n}(t)$ & standard (non MEC) traffic at time $t$\\
$\theta_{0}$ &  BS load independent energy consumption or operation energy\\
$f_{\rm max}$ & maximum processing rate for VM $m$\\
$\mathcal{F}$ & finite set of available processing rates for VM($m$)\\
$\theta_{m}^{\rm ov}(t)$ & energy overheads incurred when turning on/off VMs\\
$\theta_{{\rm idle}, m}(t)$ & static energy consumed by VM $m$ in the idle state\\
$\theta_{{\rm max}, m}(t)$  & maximum energy consumed by VM $m$ at maximum processing rate\\
$\gamma_m(t)$ & workload fraction to be computed by the $m$-th VM\\
$\gamma^{\max}$ & maximum computation load \mbox{per-VM} \\
$\Delta $ & maximum \mbox{per-slot} and \mbox{per-VM} allowed processing time\\
$\theta_{\rm idle}$ & energy consumption of network interfaces in idle mode\\
$\theta_{\rm data}$ & energy cost of exchanging one unit of data between the server and the BS\\
$\beta_{\rm max}$ & maximum energy buffer capacity \\
$\beta_{\rm up}, \beta_{\rm low}$ & upper and lower energy buffer thresholds \\
\hline
& {\bf \qquad\qquad\qquad\qquad Variables}\\
\hline
$\theta_{{\rm tot},n}(t)$ & total energy consumption for the communication site $n$\\
$\theta_{{\rm BS},n}(t)$ & BS $n$ energy cost at $t$ \\
$\theta_{{\rm MEC},n}(t)$ & server consumption due to computation activities\\
$\theta_{{\rm TX},n}(t)$ & data transmission energy consumption between the BS and the MEC server\\
$\zeta_{n}(t)$ & BS $n$ switching status indicator at $t$\\
$M(t)$ & number of \acp{VM} to be active in time slot $t$\\
$\theta_{\rm load}(t)$ & total wireless transmission power\\
$f_{m}(t)$ &  instantaneous processing rate\\
$\theta_{m}^{\rm op}(t)$ & energy consumption of VM $m$ operation\\
$\alpha_{m}(t)$ & load dependent factor\\
$\mu_m(t)$ & the expected processing time\\
$B_n(t)$ & the total amount of load that is served by the BS site\\
$\beta_{n}(t)$ & energy buffer level in slot $t$\\
$H_{n}(t)$ & harvested energy profile in slot $t$ \\
$Q_{n}(t)$ & purchased grid energy in slot $t$ \\
\hline 
\end{tabular}
\label{tab:variables}
\end{table*}
  
\subsection{Traffic Load and Energy Consumption}
\label{sub:bsload}

Mobile traffic volume exhibits temporal and spatial diversity, and also follows a diurnal behavior~\cite{Peng}. Therefore, traffic volume at individual \acp{BS} can be estimated using historical mobile traffic datasets. In this paper, real \ac{MN} traffic load traces obtained from the Big Data Challenge organized by \ac{TIM}~\cite{bigdata2015tim} are used to emulate the computational load\footnote{In fact, the dataset is not a {\it true} representative of future applications that require processing at the edge, but contains data that is exchanged with the purpose of communication. We nevertheless use it due to the difficulties in finding open datasets containing computing requests.}. Specifically, the used data was collected in the city of Milan during the month of November 2013, and it is the result of users interaction within the \ac{TIM} MN, based on Call Detail Record (CDR) files for a day considering four BS sites representing the traffic load profiles. A CDR file consists of SMS, Calls and Internet records with timestamps.
To understand the behavior of the mobile data, we have applied the \mbox{X-means} clustering algorithm~\cite{pelleg2000x} to classify the load profiles into several categories. In our numerical results, each \ac{BS} $n=1,2,\dots,N$ is assigned a load profile ${L}_{n}(t)$, which is picked at random as one of the four clusters (each cluster represents a typical BS load profile) in Fig.~\ref{fig:trace_load}. ${L}_{n}(t)$ consists of computation workloads $\Gamma_{n}(t)$ ([MB]) and standard workloads $\Gamma^\prime_{n}(t)$ ([MB]). According to~\cite{youtube_pareto}, we assume that $80$\% of ${L}_{n}(t)$ is {\it delay sensitive} and, as such, requires processing at the edge, i.e. $\Gamma_{n}(t) = 0.8 L_n(t)$, whereas the remaining $20$\% pertains to standard flows, {\it delay tolerant} traffic, i.e., $\Gamma^\prime_{n}(t) = L_n(t) - \Gamma_n(t)$. 

The total energy consumption ([$\SI{} {\joule}$]) for the communication site $n$ at time slot $t$ is formulated as follows, inspired by~\cite{online_pimrc}, \cite{mec_lyapunov}, \cite{bspower}, \cite{sleep_control} and~\cite{Liumigration}:
\begin{equation}
	\theta_{{\rm tot},n}(t) = \theta_{{\rm BS},n}(t) + \theta_{{\rm MEC},n}(t) + \theta_{{\rm TX}, n}(t)\,,
	\label{eq:bsconsupt}
\end{equation}
\noindent where $\theta_{{\rm BS}, n}(t)$ is the \ac{BS} energy consumption term, $\theta_{{\rm MEC},n}(t)$ is the MEC server consumption term due to computation activities, and $\theta_{{\rm TX}, n}(t)$ represents the data transmission energy consumption between the BS and the MEC server.\\

\noindent \textbf{BS energy consumption:} $\theta_{{\rm BS},n}(t) = \zeta_{n}(t) \theta_{0} + \theta_{{\rm load}}(t)$,  where $\zeta_{n}(t) \in  \{\varepsilon, 1\}$ is the BS switching status indicator ($1$ for {\it active mode} and $\varepsilon$ for {\it power saving mode}), $\theta_{0}$ is a constant value (load independent), representing the operation energy which includes baseband processing, radio frequency power expenditures, etc. The constant $\varepsilon \in (0,1)$ accounts for the fact that the baseband energy consumption can be scaled down as well whenever there is no or little channel activity, into a power saving mode. $\theta_{{\rm load}}(t)$ represents the total wireless transmission (load dependent) power to meet the target transmission rate from the BS to the served user(s) and to guarantee low latency at the edge. 
Since we assume a \mbox{noise-limited} channel and the guarantee of low latency requirements at the edge, $\theta_{{\rm load}}(t)$ is obtained by using the transmission model in~\cite{mec_lyapunov} (see Eq.~(5) in this reference). Here, we neglect the imbalance of traffic volumes in uplink and downlink, and also we do not account for the switching energy cost for the \ac{BS} mode transition~\cite{sleep_control} due to the fact that future \ac{BS} functions will be virtualized~\cite{BS_virtualization}.\\

\noindent \textbf{MEC server energy consumption:} it depends on the number of \acp{VM} running in time slot $t$, named $M(t) \leq M$, and on the \ac{CPU} frequency that is allotted to each virtual machine. Specifically, \acp{VM} are instantiated on top of the physical \ac{CPU} cores, and each \ac{VM} is given a share of the host server \ac{CPU}, memory and network input/output interfaces. The \ac{CPU} is the main consumer of energy in the server~\cite{Beloglazov} due to the \mbox{VM-to-CPU} share mapping. Hence, in this work we focus on the \ac{CPU} utilization only. With \mbox{$f_{m}(t) \in [0, f_{\rm max}]$} we mean the instantaneous processing rate~\cite{quadra_cpu}, expressed in bits per second that are computed
, and $f_{\rm max}$ is the maximum processing rate for \ac{VM} $m$. In this paper, $f_{m}(t)$ is set within a finite set $\mathcal{F} = \{f_0, f_1,\dots, f_{\rm max}\}$ where $f_0 = 0$ represents zero speed of the \ac{VM} (e.g., deep sleep or shutdown). At any given time $t$, the total energy consumption of a virtualized server, with $M(t)$ running \acp{VM} is:
\begin{equation}
\theta_{{\rm MEC},n}(t) = \sum_{m = 1}^{M(t)} \left ( \theta_{m}^{\rm op}(t) + \theta_{m}^{\rm ov}(t) \right ) \, , 
\label{eq:mec_energy} 
\end{equation}
\noindent where $\theta_{m}^{\rm op}(t)$ is the energy consumption of VM $m$ operation and $\theta_{m}^{\rm ov}(t) \geq 0$ is the energy cost incurred through the turning on/off the \ac{VM}, i.e., $\theta_{m}^{\rm ov}(t) >0$ only when \ac{VM} $m$ is switched on/off and it is zero otherwise. $\theta_{m}^{\rm op}(t)$ is obtained using the linear relationship between the \ac{CPU} utilization contributed by VM $m$ and the energy consumption, from~\cite{quadra_cpu} and~\cite{vm_char} (see Eq. (4) in the second reference): 
\begin{equation}
\mbox{$\theta_{m}^{\rm op}(t) = \theta_{{\rm idle}, m}(t) + \alpha_{m}(t)(\theta_{{\max}, m}(t) - \theta_{{\rm idle}, m}(t))$} \, , 
\end{equation}
where $\theta_{{\rm idle}, m}(t)$  represents the {\it static} energy drained by VM $m$ in the idle state, and $\theta_{{\max}, m}(t)$ is the {\it maximum} energy it drains. The quantity, $\alpha_{m}(t) (\theta_{{\max}, m}(t) - \theta_{{\rm idle}, m}(t))$, represents the {\it dynamic} energy component, where \mbox{$\alpha_{m}(t) = (f_{m}(t)/f_{\max})^{2}$~\cite{llcprediction}} is a load dependent factor. Note that $\alpha_m(t)$ and $f_m(t)$ are deterministically related as $f_{\max}$ is a constant. $\theta_{m}^{\rm ov}(t)$ is obtained from~\cite{vm_char} (see Eq. (5) in this reference) as a constant and is typically limited to a few hundreds of $\SI{} {\milli\joule}$ per $\SI{} {\mega\hertz}^{2}$.

Conventionally, for each \ac{BS} site, the hypervisor, i.e., the software that provides the environment in which the \acp{VM} operate, is in charge of allocating $f_{m}(t)$ and the workload fraction to be computed by the $m$-th \ac{VM}, named $\gamma_{m}(t)$. In our setup, we have $\sum_{m=1}^{M(t)} \gamma_m(t) \leq \Gamma_n(t)$, where equality is achieved when the workload is fully served by the $M(t)$ VMs. 
We also note that, in practical application scenarios, the maximum \mbox{per-VM} computation load to be computed is generally limited up to an assigned value, named $\gamma^{\rm max}$. Motivated by the energy efficient requirements from~\cite{eerequirements}, i.e., the hypervisor's ability to accept and implement policies from a management entity, in this paper, the {\it edge controller} usage is pursued. Here, the edge controller determines the $f_{m}(t)$ value that will yield the desired or expected processing time, $\mu_{m}(t) = \gamma_{m}(t)/f_{m}(t)$, considering the workload $\gamma_{m}(t)$ allotted to \ac{VM} $m$. $\mu_{m}(t)$ must be less than or equal to the maximum \mbox{per-slot} and \mbox{per-VM} processing time (in seconds), named $\Delta$, i.e., $\mu_{m}(t) \leq \Delta$. Note that $\Delta$ is also the server's response time, i.e., the maximum time allowed for processing the total computation load.

We remark that, as a result of the allocation procedure that is developed in this paper, for any \ac{BS} site $n$, the processing rates $f_{m}(t)$ shall be found, similar to~\cite{vm_char} (see remark 1 from this reference). Then, the total amount of load that is served by the \ac{BS} site may be set as: $B_n(t) = \sum_{m=1}^{M(t)} \gamma_{m}(t) \leq \Gamma_n(t)$. The objective of the considered optimization is to find the operating mode for the \ac{BS} (either ``on'' or ``power saving''), the number of \acp{VM} $M(t)$ that are to be allocated and, for each of them, the processing rate $f_m(t)$. In doing so: 1) the amount of delay sensitive load that is not served at the edge, \mbox{$\Gamma_n(t) - \sum_{m=1}^{M(t)} \gamma_{m}(t)$}, shall be minimized, while exploiting as much as possible the energy harvested from the solar panels, so that the mobile network will be energetically \mbox{self-sufficient}, and 2) the load is computed in a time shorter than or equal to $\Delta$. The details of the proposed optimization algorithm are provided in Section~\ref{sec:case1}.\\

\noindent \textbf{Data transmission energy consumption:} we assume that the \mbox{inter-communication} between the BS and the MEC server is \mbox{bi-directional} and symmetric. Hence, under \mbox{steady-state} operating conditions, for the communication site $n$, $\theta_{{\rm TX}, n}(t)$ is obtained as $\theta_{{\rm TX},n}(t) = \theta_{\rm idle}(t) + \theta_{\rm data}(t)\, B_n(t)$ by using the VM migration hint from~\cite{joint_migr}, where $\theta_{\rm idle}(t)$ (fixed value in $\rm{J}$) is the energy drained by the network interfaces in idle mode over a time slot $t$, $\theta_{\rm data}$ (fixed value in $\rm{J}/{\rm byte}$) is the cost of exchanging one byte of data between the MEC server and the BS per time slot $t$, and $B_n(t)$ is the amount of data exchanged. These parameters, $\theta_{\rm idle}(t)$ and $\theta_{\rm data}(t)$, are obtained from~\cite{joint_migr}. Note that $B_n(t)$ also corresponds to the amount of data to be processed at the MEC server in bytes.

\subsection{Energy Patterns and Storage}
\label{sub:eebuffer}

\begin{figure}[t]
	\centering
	\resizebox{\columnwidth}{!}{\input{./energy_profiles.tex}}
	\caption{Example traces for harvested solar energy from~\cite{amerinia}.}
	\label{fig:energy_trace}
\end{figure}
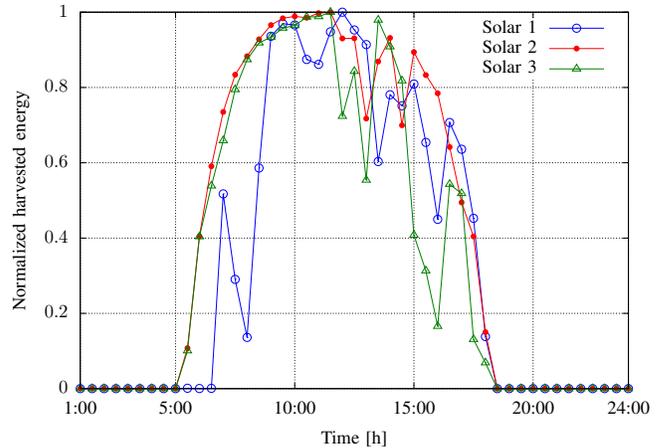 

The energy buffer is characterized by its maximum energy storage capacity $\beta_{\rm max}$. At the {\it beginning} of each time slot $t$, the EM provides the energy level report to the edge controller through the local MEC server, thus the \ac{EB} level $\beta_{n}(t)$ is known, enabling the provision of the required computation resources, i.e., the \acp{VM}. The energy level report/file from the EM to the MEC server is transferred using the pull mode procedure (e.g., File Transfer Protocol)~\cite{filetransfer}. 

In this work, the amount of harvested energy $H_{n}(t)$ in time slot $t$ in the communication site $n$ is obtained from open source solar traces~\cite{amerinia} (see Fig.~\ref{fig:energy_trace}). The dataset is the result of daily environmental records. In our numerical results, $H_{n}(t)$ represents a daily solar radiation record for three different areas. From the three solar profiles, each communication site energy profile is picked at a random to represent the daily energy harvested and then scaled to fit the \ac{EB} capacity $\beta_{\rm max}$ of $\SI{490} {\kilo\joule}$. Thus, the available \ac{EB} level $\beta_{n}(t + 1)$ at the beginning of time slot $t+1$ is calculated as follows: 
\begin{equation}
\beta_{n}(t + 1) = \beta_{n}(t) + H_{n}(t) - \theta_{{\rm tot},n}(t) + Q_{n}(t),
\label{eq:offgrid}
\end{equation}
\noindent where $\beta_{n}(t)$ is the energy level in the battery at the beginning of time slot $t$, $\theta_{{\rm tot},n}(t)$ is the energy consumption of the communication site over time slot $t$, see Eq.~(\ref{eq:bsconsupt}), and $Q_{n}(t) \geq 0$ is the amount of energy purchased from the power grid. We remark that $\beta_{n}(t)$ is updated at the beginning of time slot $t$ whereas $H_{n}(t)$ and $\theta_{{\rm tot},n}(t)$ are only known at the end of it.

For decision making in the edge controller, the received \ac{EB} level reports are compared with the following thresholds: $\beta_{\rm low}$ and $\beta_{\rm up}$, respectively termed the lower and the upper energy threshold with $0 < \beta_{\rm low} < \beta_{\rm up} < \beta_{\rm max}$. $\beta_{\rm up}$ corresponds to the desired energy buffer level at the BS and $\beta_{\rm low}$ is the lowest EB level that any \ac{BS} should ever reach. If $\beta_{n}(t) < \beta_{\rm low}$, then BS $n$ is said to be {\it energy deficient}, our optimization in the following section makes sure that $\beta_n(t)$ never falls below $\beta_{\rm low}$ due to its transmission and computing activities within a time slot. Instead, if for any time slot we have  $\beta_n(t) < \beta_{\rm up}$, then the following amount of energy $Q_n(t) = \beta_{\rm up} - \beta_n(t)$ is purchased from the energy grid to compensate for the deviation from the desired EB level (due to previous BS activity). 

\section{Optimization for a Single Communication Site} 
\label{sec:case1}

In this section, we formulate an optimization problem to obtain {\it energy savings} through \mbox{short-term} traffic load, harvested energy predictions, along with energy management procedures for a {\it single} communication site. The optimization problem is defined in section~\ref{sub:prob1}, and the communication site management procedures are presented in section~\ref{sub:site_manager}.

\subsection{Problem Formulation}
\label{sub:prob1}

At the beginning of each time slot $t$, the edge controller receives the energy level report $\beta_{n}(t)$ from each EM (via the MEC application responsible for energy profiles in the MEC server), using the pull mode file transfer. Here, we aim at minimizing the overall energy consumption in the communication site over time, i.e., the consumption related to the \ac{BS} transmission activity and the MEC server, by applying BS power saving modes and \ac{VM} \mbox{soft-scaling}, i.e., tuning the number of active virtual machines. To achieve this, we first consider the optimization for a single communication site. We define two cost functions as: 
\begin{itemize}
\item[\bf F1)] $\theta_{{\rm tot},n}(t)$, which weighs the energy consumption due to transmission (BS) and computation (MEC server); and 
\item[\bf F2)] a quadratic term $(\Gamma_n(t)-B_n(t))^2$, which accounts for the \ac{QoS} cost. 
\end{itemize}
In fact, F1 tends to push the system towards self-sustainability solutions, i.e., $\zeta_{n}(t) \to \varepsilon$. Instead, F2 favors solutions where the delay sensitive load is entirely processed by the local MEC server, i.e., $B_n(t) \to \Gamma_n(t)$. A weight $\eta \in [0,1]$, is utilized to balance the two objectives F1 and F2. The corresponding (weighted) cost function is defined as:
\begin{equation}
\label{eq:Jfunc}
J(\zeta,\alpha,t) \stackrel{\Delta}{=} \overline{\eta}\theta_{{\rm tot},n}(\zeta_n(t),\{\alpha_m(t)\},t) + \eta(\Gamma_n(t) - B_n(t))^2 \, ,
\end{equation}
where $\overline{\eta } \stackrel{\Delta}{=} 1 - \eta$, with $\{\alpha_m(t)\}$ we mean the sequence of factors $\alpha_1(1), \alpha_2(1), \dots, \alpha_{M(t)}(1)$. Hence, letting $1$ be the current time slot and $T$ be the time horizon, the following optimization problem is formulated over time slots $1,\dots,T$:
\begin{eqnarray}
        \label{eq:objt}
        \textbf{P1} & : & \min_{\bm \zeta, \bm \alpha} \sum_{t=1}^T J(\zeta,\alpha,t)  \\
        && \hspace{-1.25cm}\mbox{subject to:} \nonumber \\
        {\rm C1} & : & \zeta_{n}(t) \in \{\varepsilon,1\}, \nonumber \\
        {\rm C2} & : & b \leq M(t) \leq M, \nonumber \\
        {\rm C3} & : & \beta_{n}(t) \geq \beta_{\rm low} , \nonumber \\ 
        {\rm C4} & : & 0 \leq f_{m}(t) \leq f_{\rm max}, \nonumber \\
        {\rm C5} & : & 0 \leq \gamma_{m}(t) \leq \gamma^{\rm max}, \nonumber \\
        {\rm C6} & : & \mu_{m}(t) \leq \Delta, \quad t=1,\dots, T \, ,\nonumber    
\end{eqnarray}
\noindent where $m=1,\dots,M(t)$ (VM index), vectors $\bm \zeta$ (BS switching status in time slots $1,\dots,T$) and $\bm \alpha$ (load dependent factor) contain the \textbf{control actions} for the considered time horizon, per communication site, i.e., $\bm \zeta = [\zeta(1), \zeta(2), \dots, \zeta(T)]$ and \mbox{$\bm \alpha = [\{\alpha_m(1)\}, \{\alpha_m(2)\}, \dots, \{\alpha_m(T)\}]$}. Constraint C1 specifies the BS operation status (either {\it power saving} or {\it active}), C2 forces the required number of \acp{VM}, $M(t)$, to be always greater than or equal to a minimum number \mbox{$b \geq 1$}: the purpose of this is to be always able to handle mission critical communications. C3 makes sure that the \ac{EB} level is always above or equal to a preset threshold $\beta_{\rm low}$, to guarantee {\it energy \mbox{self-sustainability}} over time. Note that this constraint may imply that in certain time slots the BS is to be switched off, although the workload may be \mbox{non-negligible}. When managing a single BS site (the formulation in this section), this implies that the load will not be served, but this fact may be compensated for when multiple communication sites are jointly managed, e.g., handing off the workload to another, energy richer, BS. This is dealt with in Section~\ref{sec:case2}. Furthermore, C4 and C5, bound the maximum processing rate and workloads of each running VM $m$, with $m = 1,\dots, M(t)$, respectively. Constraint C6 represents a \mbox{hard-limit} on the corresponding \mbox{per-slot} and \mbox{per-VM} processing time. 

To solve {\rm P1} in Eq.~\eq{eq:objt}, we leverage the use of \ac{LLC}~\cite{llcprediction}~\cite{chung1992limited} and heuristics, obtaining the controls $\varsigma(t)\stackrel{\Delta}{=}(\zeta(t), \{\alpha(t)\})$ for $t=1,\dots,T$. Note that Eq.~\eq{eq:objt} can iteratively be solved at any time slot $t \geq 1$, by just redefining the time horizon as $t^\prime = t, t+1, \dots, t+T-1$.

\subsection{Communication Site Management}
\label{sub:site_manager}

In this subsection, a traffic load and energy harvesting prediction method, and an online management algorithm are proposed to solve the previously stated problem {\rm P1}. In subsection~\ref{predict}, we discuss the prediction of the future (\mbox{short-term}) traffic load and harvested energy processes, and then in subsection~\ref{online_proc}, we solve {\rm P1} by first constructing the \mbox{state-space} behavior of the control system, where online control key concepts are introduced. Finally, the algorithm for managing the single communication site is presented in subsection~\ref{alg}.

\subsubsection{Traffic load and energy forecasting}
\label{predict}

\ac{ML} techniques constitute a promising solution for network management and energy savings in cellular networks~\cite{machine_learning_tut}\cite{machine_learning_details}. In this work, given a time slot duration of $\tau =\SI{30} {\minute}$, we perform time series prediction, i.e., we obtain the $T = 3$ estimates of $\hat{L}_{n}(t)$ and $\hat{H}_{n}(t)$, by using an \ac{LSTM} network developed in Python using Keras deep learning libraries (Sequential, Dense, LSTM) where the network has a visible layer with one input, one hidden layer of four \ac{LSTM} blocks or neurons, and an output layer that makes a single value prediction. This type of recurrent neural network uses \mbox{back-propagation} through time for learning and memory blocks for regression~\cite{lstmlearn}. The dataset is split as $67\%$ for training and $33\%$ for testing. The network is trained using $100$ epochs ($2,600$ individual training trials) with batch size of one. As for the performance measure of the model, we use the Root Mean Square Error (RMSE). The prediction steps are outlined in Table~\ref{lstm_model}. Fig.~\ref{fig:bs_load} and Fig.~\ref{fig:energy_load} show the prediction results that will be discussed in Section~\ref{sec:results}.

\begin{table}[tp]
    \caption{LSTM Prediction Model Steps}
    \center
    \resizebox{\columnwidth}{!}{%
    \begin{tabular} {|l|}
          \hline 
          {\bf Modeling steps} \\ 
          \hline
          Step 1: load and normalize the dataset\\
          Step 2: split dataset into training and testing\\
          Step 3: reshape input to be [samples, time steps, features]\\
          Step 4: create and fit the LSTM network\\
          Step 5: make predictions\\
          Step 6: calculate performance measure\\
          \hline 
     \end{tabular}%
     }
     \label{lstm_model}
\end{table}

\subsubsection{Edge system dynamics}
\label{online_proc}

we denote the system state vector at time $t$ by $\bm{x}(t) = (M(t),\beta_{n}(t))$, which contains the number of active VMs, $M(t)$, and the EB level, $\beta_n(t)$, for the BS site $n$. \mbox{$\bm{\varsigma}(t)=(\zeta(t),\{\alpha_m(t)\})$} is the input vector, i.e., the control action that drives the system behavior at time $t$. The system evolution is described through a \mbox{discrete-time} \mbox{state-space} equation, adopting the \ac{LLC} principles~\cite{llcprediction}~\cite{chung1992limited}:
\begin{equation}
\bm{x}(t + 1) = \Phi(\bm{x}(t),\bm{\varsigma}(t)) \, , 
\end{equation}
\noindent where  $\Phi(\cdot)$ is a behavior model that captures the relationship between $(\bm{x}(t),\bm{\varsigma}(t))$, and the next state $\bm{x}(t + 1)$. Note that this relationship accounts for 1) the amount of energy drained $\theta_{{\rm tot},n}(t)$, that harvested $H_{n}(t)$ and that purchased from the power grid $Q_n(t)$, which together lead to the next buffer level $\beta_{n}(t+1)$ through Eq.~\eq{eq:offgrid}, and 2) to the traffic load $L_{n}(t)$, from which we compute the server workloads $\Gamma_{n}(t)$, that leads to $M(t)$ and to the control $\bm{\varsigma}(t)$. The network management algorithm in the edge controller, the \ac{ENAAM} algorithm, finds the best control action vector for the communication site, following a {\it model predictive control approach}. Specifically, for each time slot $t$, problem \eq{eq:objt} is solved, obtaining control actions for the whole time horizon $t,t+1,\dots,t+T-1$. The control action that is applied at time $t$ is $\bm{\varsigma}^{*}(t)$, which is the first one in the retrieved control sequence. This control amounts to setting the BS radio mode according to $\zeta^*(t)$, i.e., either active or power saving, and the number of instantiated \acp{VM}, $M^*(t)$, along with their obtained $\{\alpha_m^{*}(t)\}$ values (see remarks 1 and 2 below). This is repeated for the following time slots $t+1, t+2, \dots$.\\

\subsubsection*{\bf Remark 1 (Role of prediction)} State $\bm{x}(t)$ and control $\bm{\varsigma}(t)$ are respectively measured and applied at the beginning of time slot $t$, whereas the offered load $L_{n}(t)$ and the harvested energy $H_{n}(t)$ are accumulated during the time slot and their value becomes known only by the end of it. This means that, being at the beginning of time slot $t$, the system state at the next time slot $t+1$ can only be estimated, which we formally write as:
\begin{equation}
       \hat{\bm{x}}(t + 1) = \Phi(\bm{x}(t),\bm{\varsigma}(t)) \, ,
       \label{eq:state_forecast}
\end{equation}
the same applies to the subsequent time slots in the optimization horizon $t+2, t+3, \dots, t+T-1$. For these estimations we use the forecast values of load $\hat{L}_{n}(t)$ and harvested energy $\hat{H}_{n}(t)$, from the LSTM forecasting module.\\

\subsubsection*{\bf Remark 2 (VM number and workload allocation)} a remark on the provisioned VMs per time slot \mbox{per-MEC} server, $M(t)$, is in order. Specifically, the number of active \ac{VM} (i.e., the VM computing cluster) depends on the predicted load, $\hat{L}_{n}(t+1)$, where the expected server workload is \mbox{$\hat{\Gamma}_{n}(t+1) = 0.8 \hat{L}_{n}(t+1)$}. Each \ac{VM} can compute an amount of up to $\gamma^{\max}$. Then, an estimate of the number of virtual machines that shall be active in time slot $t$ to serve the predicted server workloads is here obtained as: \mbox{$M(t) = \ceil[\big] {(\hat{\Gamma}_{n}(t+1)/ \gamma^{\max})}$}, where $\ceil[\big] {\cdot}$ returns the nearest upper integer. We heuristically split the workload among virtual machines by allocating a workload $\gamma_m(t)=\gamma^{\max}$ to the first $M(t)-1$ \acp{VM}, $m=1,\dots,M(t)-1$, and the remaining workload $\gamma_{m}(t) = \hat{L}_{n}(t+1) - (M(t)-1)\gamma^{\max}$ to the last one $m=M(t)$.\\

\noindent\textbf{Controller decision-making:} the controller is obtained by estimating the relevant parameters of the operating environment, i.e., the BS load $\hat{L}_n(t)$ and the harvested energy $\hat{H}_n(t)$, and subsequently using them to forecast the future system behavior through Eq.~\eq{eq:state_forecast} over a \mbox{look-ahead} time horizon of $T$ time slots. 
The control actions are picked by minimizing $J(\zeta,\alpha,t)$, see Eq.~\eq{eq:Jfunc}. At the beginning of each time slot $t$ the following process is iterated: 
\begin{itemize} 
\item[\bf 1)] Future system states, $\hat{\bm{x}}(t+k)$, for a prediction horizon of $k = 1, \dots, T$ steps are estimated using Eq.~\eq{eq:state_forecast}. These predictions depend on past inputs and outputs up to time $t$, on the estimated load $\hat{L}_{n}(\cdot)$ and energy harvesting $\hat{H}_{n}(\cdot)$ processes, and on the control $\bm{\varsigma}(t+k)$, with $k =  0, \dots, T-1$.
\item[\bf 2)] The sequence of controls $\{\bm{\varsigma}(t+k)\}_{k=0}^{T-1}$ is obtained for each step of the prediction horizon by optimizing the weighted cost function $J(\cdot)$, see Eq.~\eq{eq:Jfunc}.
\item[\bf 3)] The control $\bm{\varsigma}^*(t)$ corresponding to the first control action in the sequence with the minimum total cost is the applied control for time $t$ and the other controls $\bm{\varsigma}^*(t+k)$ with $k = 1, \dots, T-1$ are discarded. 
\item[\bf 4)] At the beginning of the next time slot $t+1$, the system state $\bm{x}(t+1)$ becomes known and the previous steps are repeated. 
\end{itemize}

\subsubsection{The ENAAM algorithm}
\label{alg}

Let $t$ be the current time. $\hat{L}_{n}(t+k-1)$ is the forecast load in slot $t+k-1$, with $k=1,\dots,T$, i.e., over the prediction horizon. For the control to be feasible, we need $\underline{\Gamma}_{n}(t) \leq B_{n}(t) \leq \hat{\Gamma}_{n}(t+k-1)$, where $\underline{\Gamma}_{n}(t)$ is the smallest $\Gamma$ such that $\texttt{round}(\hat{\underline{\Gamma}}_{n}(t+1)/ \gamma^{\max}) = b$. For the buffer state, we heuristically set $\zeta(t+k-1) = \varepsilon$ if either $\beta_{n}(t+k-1) < \beta_{\rm low}$ or $L_{n}(t+k-1) < L_{\rm low}$, and $\zeta(t+k-1) = 1$ otherwise ($\beta_{\rm low}$ and $L_{\rm low}$ are preset low thresholds for the \ac{EB} and the \ac{BS} load, respectively). For slot $t+k-1$, the feasibility set $\mathcal A (t+k-1)$ contains the control pairs $(\zeta(t),\{\alpha_m(t)\})$ that obey these relations. 

\begin{small}
\begin{algorithm}[t]
\begin{tabular}{l l}
{\bf Input:}  & $\bm{x}(t)$ (current state) \\
{\bf Output:} & $\bm{\varsigma}^{*}(t) = (\zeta^*(t),\{\alpha_m^*(t)\})$ \\
01:		& \hspace{-1cm}Initialization of variables\\
		& \hspace{-1cm}${\mathcal S}(t) = \{\bm{x}(t)\}$, ${\rm Cost}(\bm{x}(t))=0$ \\
02:		& \hspace{-1cm}{\bf for} $k = 1, \dots, T$ {\bf do}\\
		& \hspace{-1cm}\quad - forecast the load $\hat{L}_{n}(t+k-1)$ \\
		&\hspace{-1cm}\quad - forecast the harvested energy $\hat{\rm H}_{n}(t+k-1)$ \\
		& \hspace{-1cm}\quad - ${\mathcal S}(t+k) = \emptyset$ \\
03:		& \hspace{-1cm}\quad {\bf for all} $\bm{x} \in {\mathcal S}(t+k-1)$ {\bf do}\\
04:		& \hspace{-1cm}\qquad {\bf for all} $\bm{\varsigma} = (\zeta,\{\alpha_m(t)\}) \in {\mathcal A}(t+k-1)$ {\bf do}\\
05:		& \hspace{-1.1cm}\quad\quad\quad $\hat{\bm{x}}(t+k) = \Phi(\bm{x}(t+k-1),\bm{\varsigma})$\\
06:		& \hspace{-1.1cm}\quad\quad\quad ${\rm Cost}(\hat{\bm{x}}(t+k)) =  J(\zeta, \alpha, t+k-1)$\\
		& \hspace{1.25cm} $ + {\rm Cost}(\bm{x}(t+k-1),\bm{\varsigma})$\\
07:		& \hspace{-1.1cm}\quad\quad\quad ${\mathcal S}(t+k) = {\mathcal S}(t+k) \cup \{\hat{\bm{x}}(t+k)\}$\\
		& \hspace{-1cm}\qquad {\bf end for}\\
		& \hspace{-1cm}\quad {\bf end for}\\
		& \hspace{-1cm}{\bf end for}\\
08:		& \hspace{-1cm}{\bf Find $\hat{\bm{x}}_{\min} = {\rm argmin}_{\hat{\bm{x}} \in {\mathcal S}(t+T)} {\rm Cost}(\hat{\bm{x}}) $}\\
09:		& \hspace{-1cm}{$\bm{\varsigma}^{*}(t) :=$ control leading from $\bm{x}(t)$ to $\hat{\bm{x}}_{\min}$}\\
10:		& \hspace{-1cm}{\bf Return $\bm{\varsigma}^{*}(t)$}
\end{tabular}
\caption{ENAAM}
\label{algo:enaam}
\end{algorithm}
\end{small}

The algorithm is specified in Alg.~\ref{algo:enaam} as it uses the technique in~\cite{llcprediction}: the search starts (line 01) from the system state at time $t$, $\bm{x}(t)$, and continues in a \mbox{breadth-first} fashion, building a tree of all possible future states up to the prediction depth $T$. A cost is initialized to zero (line 01) and is accumulated as the algorithm travels through the tree (line 06), accounting for predictions, past outputs and controls. The set of states reached at every prediction depth $t+k$ is referred to as $\mathcal S(t+k)$. For every prediction depth $t+k$, the search continues from the set of states $\mathcal S(t+k-1)$ reached at the previous step $t+k-1$ (line 03), exploring all feasible controls (line 04), obtaining the next system state from Eq.~\eq{eq:state_forecast} (line 05), updating the accumulated cost as the result of the previous accumulated cost, plus the cost associated with the current step (line 06), and updating the set of states reached at step $t+k$ (line 07). When the exploration finishes, the initial action (at time $t$) that leads to the best final accumulated cost, at time $t+T-1$, is selected as the optimal control $\bm{\varsigma}^*(t)$ (lines 08, 09, 10). Finally, for line 04, we note that $\Gamma_{n}$ belongs to the continuous set $[\underline{\Gamma}_{n}, \hat{L}_{n}(t+k-1)]$. To implement this search, we quantized this interval into a number of equally spaced points, obtaining a search over a finite set of controls.  \\

\noindent\textbf{ENAAM complexity:} the computation complexity of the algorithm is $O(N_{x} N_{\varsigma} T)$, where $N_x \stackrel{\Delta}{=} |\bm{x}(t)|$ and $N_{\varsigma} \stackrel{\Delta}{=} |\bm{\varsigma}(t)|$ respectively represent the number of system states and the number of feasible actions at time $t$. Note that state and action space are respectively quantized into $N_x = M \times N_\beta$ and $N_\varsigma = 2 \times M \times N_\alpha$ levels, where $M$ is the number of virtual machines, $N_\beta$ is the number of quantization levels for the energy buffer and $N_\alpha$ is the number of quantization levels for the load variable $\alpha_m(t)$. Such quantization facilitates the search in Alg.~\ref{algo:enaam}. Note that exhaustive search would entail a complexity of $O((N_{x} N_{\varsigma})^T)$.

\section{Multiple Communication Sites}
\label{sec:case2}

In this section, we extend the work from section~\ref{sec:case1} by considering the energy savings for {\it multiple} communication sites. We formulate an optimization problem to obtain energy savings through \mbox{short-term} traffic load and harvested energy predictions, clustering, along with energy management procedures for the clustered BS sites. The problem formulation for multiple communication sites is described in section~\ref{sub:prob2}, then cluster formation is discussed in section~\ref{sub:cluster}, and the edge management procedure for each cluster, enabled by the edge controller, is presented in section~\ref{sub:site2_manager}.

\subsection{Problem Formulation}
\label{sub:prob2}

Our objective is to improve the overall energy savings of the network by clustering BSs based on their location (or distance measures) similarity, and then optimizing the energy savings within each cluster by employing the single optimization case described in section~\ref{sec:case1}. From an energy efficiency perspective, in a cluster of \ac{BS} nodes, one \ac{BS} (or more) might have a preference of switching off, by first offloading its (their) traffic load to its (their) neighboring BS that have enough spare capacity for handling extra traffic load, and then switching off. The whole offloaded traffic load from the BS, denoted by BS $n$, is allocated to  the neighboring cluster member (active BS) in which orthogonal resource allocation helps mitigate \mbox{intra-cluster} interference, such that the selected neighboring BS, denoted by BS $n^\prime$, is allocated the incremental load, denoted by $L_{nn^\prime}(t) \stackrel{\Delta}{=} L_n(t)$. Whenever a BS is switched off, it should maintain service to its users via a \mbox{re-association} process in order to offload the users to the neighboring active \ac{BS} having extra resources for handling upcoming extra traffic load. The \mbox{re-association} process involves notifying the connected users to try and connect to neighboring \acp{BS} with extra resources. 

In the view of the above, we consider that all BSs are grouped into sets of clusters $\mathcal{O} = \{O_1, \dots, O_{|\mathcal{O}|}\}$. Here, a given cluster $O_i \in \mathcal{O}$, with $i=1,\dots, |\mathcal{O}|$, consists of a set of \acp{BS} that coordinate with the controller. The clustering mechanism is discussed in Section~\ref{sub:cluster}. For each cluster $O_i \in \mathcal{O}$, we aim to minimize the energy consumption, i.e., the consumption due to \ac{BS} transmission and the running \acp{VM} in the servers, using BS power saving modes and \ac{VM} \mbox{soft-scaling} per active cluster member. To do so, we define a cost function which captures the individual communication site energy consumption and its \ac{QoS}. The (weighted) cost for each cluster member, BS $n \in O_i$, is redefined as:
\begin{equation}
	\label{eq:J2func}
	J_n(\zeta,\alpha,t) \stackrel{\Delta}{=} \overline{\eta}\theta_{{\rm tot},n}(\zeta_n(t),\{\alpha_m(t)\}_n,t) + \eta(\Lambda_n(t) - B_n(t))^2 \, ,
\end{equation}
where $\zeta_n(t)$ is the activity status of BS $n$ (either {\it power saving} or {\it active}), $\{\alpha_m(t)\}_n$ is the set of factors for the allocated VMs at BS $n$. Moreover, $\Lambda_n(t) \leftarrow L_n(t)$ if BS $n$ only handles its own traffic, whereas \mbox{$\Lambda_n(t) \leftarrow L_n(t)+ \Delta L_{n}(t)$}, in case one (or multiple) BSs are switched off in time slot $t$ and its (their) traffic is redirected (handed off) to BS $n$. The computation of $\Delta L_{n}(t)$ is addressed in section~\ref{sub:site2_manager}. The per cluster cost $\Upsilon_{O_i}(\bm\zeta_i,\bm\alpha_i,t)$ is the aggregated cost of all cluster members, $\Upsilon_{O_i}(\bm\zeta_i,\bm\alpha_i,t) = \sum_{\forall n \in O_i} J_n(\zeta,\alpha,t)$. Hence, over time horizon, $t=1,\dots,T$, the following optimization problem is defined:
\begin{eqnarray}
	\label{eq:cluster_objt}
	\textbf{P2} & : & \min_{\mathcal{\bm E}} \sum_{\forall O_i \in \mathcal{O}} \Upsilon_{O_i}(\bm\zeta_i,\bm\alpha_i,t) \\
	&& \hspace{-1.25cm}\mbox{subject to:} \nonumber \\
        {\rm C1-C6} & : & \textrm{ from Eq.~\eq{eq:objt}},  \nonumber \\
        {\rm C7} & : & |O_i| \geq 1, \forall \, O_i \in \mathcal{O}, \nonumber \\
        {\rm C8} & : & O_i \cap O_j = \emptyset, \forall \, O_i, O_j \in \mathcal{O}, O_i \neq O_j \nonumber,
\end{eqnarray}
\noindent where $\mathcal{E} \stackrel{\Delta}{=} \{\bm \zeta_i, \bm \alpha_i\}$ is the collection of variables to be reconfigured for all the BS clusters (the whole MN), for all time slots $t=1,\dots,T$. As for the constraints, C7 and C8 ensure that each BS is part of only one cluster. Solving {\bf P2} in Eq.~\eq{eq:cluster_objt} involves BS clustering, the forecasting method from section~\ref{predict}, a heuristic rule for the selection of which BSs have to be switched off, and the ENAAM algorithm from section~\ref{alg}. Once {\bf P2} is solved, the control action to be applied at time $t$, per cluster $O_i$, corresponds to the elements in $\{\bm \zeta_i, \bm \alpha_i\}$ that are associated with the first time slot $1$ in the optimization horizon. As above, Eq.~\eq{eq:cluster_objt} can iteratively be solved at any time slot $t \geq 1$, by just redefining the time horizon as $t^\prime = t, t+1, \dots, t+T-1$.

\subsection{Cluster Formation}
\label{sub:cluster}

Clustering algorithms have been proposed as a way of enabling energy saving mechanisms in \acp{BS}, where groups of inactive \acp{BS} or \acp{BS} with low loads are switched off. With the advent of EH BSs, the \acp{BS} with $\beta_{n}(t) < \beta_{\rm low}$ can be switched off, while still guaranteeing the \ac{QoS} through the other active BSs. That is, within each formed cluster, the controller tries to minimize the cost function, which captures the \mbox{trade-off} between the energy efficiency and the \mbox{QoS} of each cluster member. The key step in clustering is to identify similarities or distance measures between \acp{BS} in order to group \acp{BS} with similar characteristics. In this paper, we use the location of the \acp{BS} as it defines the relative neighborhood (the distance measures) with the other \acp{BS}. Using the location of the \acp{BS} and the distance between the \acp{BS}, we obtain a \mbox{distance-based} similarity matrix $\bm W^d$. In addition, we assume that the network topology is static during the clustering algorithm execution.

In the next section~\ref{BS_adj} we detail the clustering measure that we use to obtain the similarities between \acp{BS} based on location, followed by the \mbox{distance-based} clustering algorithm in section~\ref{cluster_alg}.

\subsubsection{Relative neighborhood based on BS adjacency and Gaussian similarity}
\label{BS_adj}

similar to~\cite{samarakoon2016dynamic}, we model the MN as a graph $G = (\mathcal{N}, E)$, where $\mathcal{N}$ represents the set of \acp{BS}, while the set $E$ contains the edges between any two BSs. There is an edge $(n,n^\prime) \in E$ if and only if $n$ and $n^\prime$ can mutually receive each other's transmission. In this case, we say that $n$ and $n^\prime$ are neighbors. We use a parameter $r_{nn^\prime}$ to characterize the presence of a link between nodes, where $r_{nn^\prime} \in \{0,1\}$. Let $y_n$ be the coordinates of BS $n \in \mathcal{N}$ in the Euclidean space. The relative neighborhood of BS $n$ is defined by the nearness of the BSs in its \mbox{$e_d$-radio} propagation space (or neighborhood):
\begin{equation}
\mathcal{Z}_n = \{n^\prime \, \textrm{s.t.} \, \left\Vert y_n -y_{n^\prime} \right\Vert \leq e_d \}.
\label{eq:rel_adj}
\end{equation}
If $n^\prime \in \mathcal{Z}_n$ we say that BSs $n$ and $n^\prime$ are neighbors, and we set $r_{nn^\prime} = 1$, otherwise $r_{nn^\prime} = 0$. The links between the vertices in $\mathcal{N}$ are weighted based on their similarities. Based on the distance between BS $n$ and $n^\prime$, we can classify the BSs based on their location using the Gaussian similarity measure~\cite{samarakoon2016dynamic} (a classification kernel function used in machine learning), which is defined as: 
\begin{equation}
        w_{nn^\prime}^d =
        \left\{ \begin{array}{ll}
            \displaystyle \text{exp} \left(\frac{-\left\Vert y_n -y_{n^\prime} \right\Vert^2}{2\sigma_d^2}\right) & \text{if} \quad \left\Vert y_n -y_{n^\prime} \right\Vert \leq e_d,\\
            0 &\text{otherwise},
        \end{array} \right.
        \label{eq:gauss_sim}
\end{equation}
where $2\sigma_d^2$ adjust the impact of the neighborhood size. In Eq.~\eq{eq:gauss_sim}, we assume that the \acp{BS} located far from each other have low similarities, compared to those that are close to each other, as those that are close are more likely to cooperate with each other. The \mbox{distance-based} similarity matrix $\bm W^d$ is formed using $w_{nn^\prime}^d$ as the ($n,n^\prime$)-th entry.

\subsubsection{Distance-based clustering}
\label{cluster_alg}

the BS clustering is performed after obtaining the similarity matrix $\bm W^d$ of the MN graph \mbox{$G = (\mathcal{N}, E)$}. Given the matrix $\bm W^d$, we employ a centralized clustering method, specifically the \mbox{K-means}~\cite{k_means}, as the matrix provides the full location knowledge. \mbox{K-means} partitions the set of nodes into clusters in which each node belongs to the cluster with the nearest mean distance. In addition, the value of $K$, i.e., the number of clusters ($|O_i|$), is known prior and is a design parameter. This algorithm requires knowledge of all the BS locations, thus, it is categorized as a centralized method. In our case, this process does not incur any computation delay as the edge controller is assumed to have high computation capabilities.



\subsection{Edge Network Management}
\label{sub:site2_manager}

Our aim is to implement and validate an \ac{LLC} framework for dynamic resource provisioning in multiple communication sites with the goal of achieving energy savings within the access network through \ac{BS} sleep modes and \ac{VM} \mbox{soft-scaling}. Given the formation of clusters, load and energy forecasting, our next goal is to developed a mechanism for solving {\bf P2} (Eq.~(\ref{eq:cluster_objt})) where each cluster of BSs adjust its transmission parameters and its computing cluster entities based on the forecast information. In order to minimize the per cluster cost function, we introduce the notion of {\it network impact} in Section~\ref{impact}, whereas we describe the edge management procedure in Section~\ref{alg2}.

\subsubsection{Network Impact}
\label{impact}

The dynamic \ac{BS} switching off strategies may have an impact on the network due to the traffic load that is offloaded to the neighboring BSs. To avoid this, the BS to be switched off must be carefully identified within a \ac{BS} cluster. To determine whether a particular \ac{BS} can be switched off or not, we follow the work done in~\cite{E_oh}. As an example, we consider one cluster $O_i$, together with its cluster members $ n \in O_i$, then from it we choose one BS, BS $n$, where BS $n$ neighbors set is denoted by $\mathcal{N}_n$. Note that the BS $n^\prime \in \mathcal{N}_n$ is the BS to which the traffic load will be offloaded to after turning off BS $n$. Also, BS $n$ can only be switched off if there exists a neighboring BS $n^\prime$ that satisfies the following feasibility constraint~\cite{E_oh}: 
\begin{equation}
L_{n^\prime}(t) + L_{nn^\prime}(t) \leq 1, \quad n^\prime \in \mathcal{N}_n,
\label{eq:load_constraint}
\end{equation}
where $L_{n^\prime}(t)$ is the original BS $n^\prime$ traffic load and $L_{nn^\prime}(t)$ is the incremental traffic load from BS $n$ (the switched off BS) to BS $n^\prime$ (the neighboring BS). We recall that the load $L_{n^\prime}(t)$ is normalized with respect to the maximum load that a BS can sustain, so the inequality in Eq.~\eq{eq:load_constraint} means that it is feasible for BS $n^\prime$ to take the extra load from BS $n$.
To quantify how the incremental system load affects the overall network load due to the switching off process, we introduce the notion of {\it network impact}. For every BS $n$ within cluster $O_i$, $i=1,\dots,K$, its {\it network impact} due to the offloaded system load onto one of the neighboring BSs is defined as:
\begin{equation}
I_n(t) = \max_{n^\prime \in \mathcal{N}_n} [ L_{n^\prime}(t) + L_{nn^\prime}(t) ], \forall \, n \in O_i.
\label{eq:load_impact}
\end{equation}
Here, the maximum network impact value $I_n(t)$ over the neighboring BSs is considered as a measure for each BS towards switching off and generating extra traffic loads for its neighboring BSs. In this work, considering cluster $O_i$, we switch off the BS $n^*$ that has the least network impact, i.e.,
\begin{equation}
n^* = \arg \!\! \min_{n \in O_i} I_n(t).
\label{eq:bs_switching_cond}
\end{equation}
The BS that takes the load from $n^*$ is selected as the BS $n^\prime$ that minimizes $L_{n^\prime}(t) + L_{n^*n^\prime}(t)$ over the set of active BSs that are on within the cluster $O_i$. For BS $n^\prime$, we then set $L_{n^\prime}(t) \leftarrow L_{n^\prime}(t) + L_{n^*n^\prime}(t)$. This procedure is sequentially repeated for all the cluster members until there is no active BS whose neighbors satisfy the feasibility condition of Eq.~\eq{eq:load_constraint}. Note that here, we focus only on which BS to switch off, as for the BS turning on state, we assume that the {\it commitment time} (time configured so that the BS automatically wakes up without external triggers) is a system parameter that is \mbox{pre-configured} when the BS is switched off. 

\subsubsection{Edge management procedure}
\label{alg2}

Here, we propose a distributed edge network management procedure that makes use of the ENAAM algorithm (see section~\ref{alg}). The decision making criterion only depends on the BS information and on its neighboring BSs, thus, the BS switching off decision can be localized within each cluster. To decide which BSs shall be switched off, we follow a sequential decision process. While this is heuristic, it iallows coping with the high complexity associated with an optimal (all BSs are jointly assessed) allocation approach. The edge management procedure is as follows. 

For each BS cluster $O_i$, with $i=1,\dots,K$, do: 
\begin{itemize}
\item[{\bf 1)}] Initialize an allocation variable $\Delta L_n(t)=0$ for all BSs $n \in O_i$. Compute $I_n(t)$, using Eq.~\eq{eq:load_impact}, for all BSs $n$ and obtain the BS with the least {\it network impact} $n^*(t)$, using Eq.~\eq{eq:bs_switching_cond}. Switch off BS $n^*(t)$ and assign its load to the neighboring BS $n^\prime \in O_i$ that minimizes $L_{n^\prime}(t) + \Delta L_{n^\prime}(t) + L_{n^*n^\prime}(t)$. Update the extra allocation for BS $n^\prime$ as $\Delta L_{n^\prime}(t) \leftarrow \Delta L_{n^\prime}(t) + L_{n^*n^\prime}(t)$. Recompute $I_n(t)$ for all the BSs that are still on and identify the next BS that can be switched off, i.e., the one with the {\it least network impact}. This procedure is repeated until none of the BSs in the cluster verifies Eq.~\eq{eq:load_constraint}. At this point, we have identified all the BSs $n^*$ that shall be switched off in $O_i$. 
\item[{\bf 2)}] For each active BS $n^\prime \in O_i$, the ENAAM algorithm is executed using $L_{n^\prime}(t) + \Delta L_{n^\prime}(t)$, where $\Delta L_{n^\prime}(t)=0$ if BS $n^\prime$ does not take extra load, whereas it is greater than zero otherwise. Note that, $\Delta L_{n^\prime}(t)$ corresponds to the total traffic that is handed over to BS $n^\prime$, possibly from multiple nearby BSs. 
\end{itemize}

\noindent\textbf{Edge network management complexity:} The algorithm is independently executed for each cluster and the corresponding time complexity is obtained as follows. Considering the action {\bf Step 1}, from above, the time complexity associated with the computation of the BS having the least network impact is linear with the size of the cluster $|O_i|$. Once that is computed, the complexity associated with updating the load allocation for the active BSs is $|O_i|-1$, which leads to a total complexity of $|O_i| (|O_i|-1)=O(|O_i|^2)$. Moreover, such process is iterated for each BS that is switched off. In the worst case, where all the BSs but one are switched off, the final complexity of step 1 is $O(|O_i|^3)$. As for {\bf Step 2}, from above, the computation complexity depends on the ENAAM algorithm, which is independently executed by each {\it active} BS. Thus, in the worse case (no BSs are switched off), the total aggregated complexity is: $O(|O_i| N_x N_\varsigma T)$, which is linear in all variables, namely, number of cluster members, number of BS states, number of actions and time horizon $T$.

\section{Performance Evaluation}
\label{sec:results}

In this section, we show some selected numerical results for the scenario of Section~\ref{sec:sys}. The parameters that were used for the simulations are listed in Table~\ref{tab_opt}.

\begin{table}
	\caption{System Parameters.}
	\center
	\begin{tabular} {|l| l|l|}
		\hline 
		{\bf Parameter} & {\bf Value} \\ 
		\hline
		Total \acp{BS}, $N$ & $24$ \\
		Max. number of \acp{VM}, $M$ &  $27$\\
		Min. number of \acp{VM}, $b$ & $1$ \\
		Time slot duration, $\tau$ &  $\SI{30} {\minute}$\\
		Operating power, $\theta_{0}$ & $\SI{10.6} {\watt}$\\
		Energy overheads for switching VM, $\theta_{m}^{\rm ov}(t)$ & $\SI{0.05} {\joule / {\mega\hertz}^{2}}$\\
		Max. computation workload per VM, $\gamma^{\max}$ & $\{5, 10\}$ MB\\
		Max. allowed processing time, $\Delta$ & $\SI{0.8} {\second}$\\
		Energy cons. of network interfaces, $\theta_{\rm idle}$ & $\SI{3} {\joule}$\\
		Cost of exchanging one unit of data, $\theta_{\rm data}$ & $\SI{6} {\joule}$/byte \\  
		Processing rate set, $\mathcal{F}$  & $\{0,4,8,12,16,20\}$\\
		Static energy consumed by VM, $\theta_{{\rm idle}, m}(t)$ & $\SI{4} {\joule}$\\
        		Max. energy cons. by VM at $f_{\rm max}$, $\theta_{{\rm max}, m}(t)$  & $\SI{10} {\joule}$\\
		Energy storage capacity, $\beta_{\rm max}$ & $\SI{490} {\kilo\joule}$\\
		Lower energy threshold, $\beta_{\rm low}$  & $30$\% of $\beta_{\rm max}$\\
		Upper energy threshold, $\beta_{\rm up}$  & $70$\% of $\beta_{\rm max}$\\
		Low traffic threshold, $L_{\rm low}$ & $4$~MB \\ 
		\hline 
	\end{tabular}
	\label{tab_opt}
\end{table}
\subsection{Simulation Setup}
\label{sub:sim_setup}

We consider multiple \acp{BS}, each one \mbox{co-located} with a MEC server and a coverage radius of $\SI{40} {\meter}$. In addition, we use a virtualized server with specifications from~\cite{specs_online} for a VMware ESXi \mbox{5.1-ProLiant} DL380 Gen8. Our time slot duration $\tau$ is set to $\SI{30} {\minute}$ and the time horizon is set to $T=3$ time slots. The simulations are carried out by exploiting the Python programming language.

\begin{table}[t]
	\caption{Average prediction error (RMSE) for harvested energy and traffic load processes, both normalized in $[0,1]$.}
	\center
	\begin{tabular} {|l|l|l|l|}
		\hline 
		&{$T = 1$} & {$T = 2$} & {$T = 3$}\\ 
		\hline
		$L(t)$ & $0.037$ & $0.042$ & $0.048$ \\ \hline
		$H(t)$ & $0.011$ & $0.016$ & $0.021$\\
	    \hline 
	\end{tabular}
	\label{tab_rmse}
\end{table}

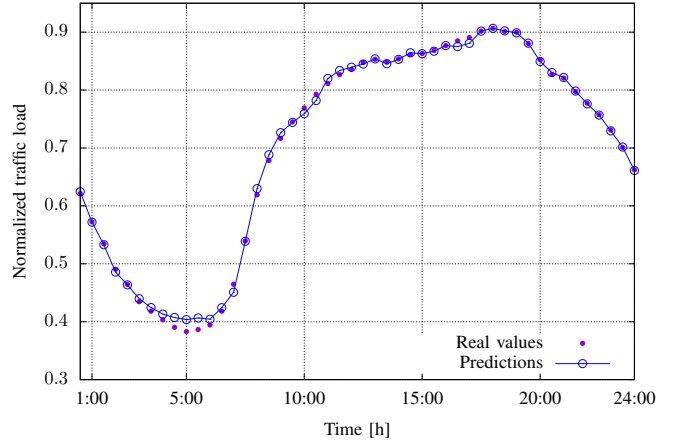
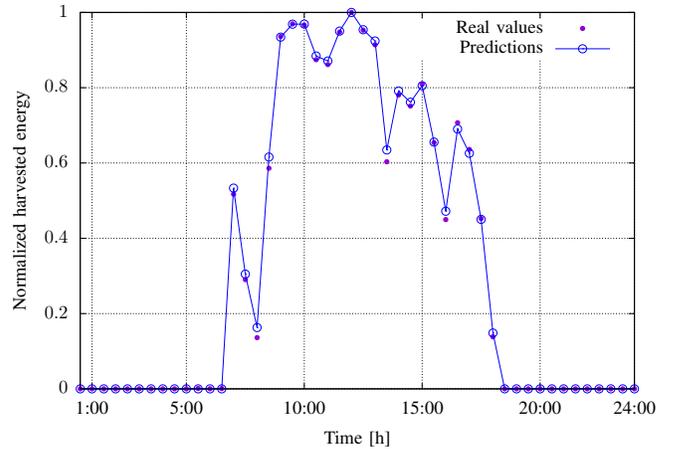
\begin{figure}[t]
	\centering
	\begin{subfigure}[t]{\columnwidth}
		\centering
		\resizebox{\columnwidth}{!}{\input{load_pred_info.tex}}
		\caption{One-step ahead predictive mean value for $L(t)$.}
		\label{fig:bs_load}	
	\end{subfigure}
	\quad
	\begin{subfigure}[t]{\columnwidth}
		\centering
		\resizebox{\columnwidth}{!}{\input{energy_pred_info.tex}}
		\caption{One-step ahead predictive mean value for $H(t)$.}
		\label{fig:energy_load}	
	\end{subfigure}
	\centering
	\caption{One-step online forecasting for both $L(t)$ and $H(t)$ patterns.}
	\label{figure:patterns}
\end{figure}

\subsection{Numerical Results}
\label{sub:num_results}

\noindent \textbf{Pattern forecasting}: we show real and predicted values for the traffic load and harvested energy over time in Figs.~\ref{fig:bs_load} and~\ref{fig:energy_load}, where we track the \mbox{one-step} predictive mean value at each step of the online forecasting routine. Then, Table~\ref{tab_rmse} shows the average RMSE of the normalized harvested energy and traffic load processes, for different time horizon values, $T \in \{1, 2, 3\}$. Note that the predictions for $H(t)$ are more accurate than those of $L(t)$ (confirmed by comparing the average RMSE), due to differences in the used dataset granularity. However, the measured accuracy is deemed good enough for the proposed optimization.\\

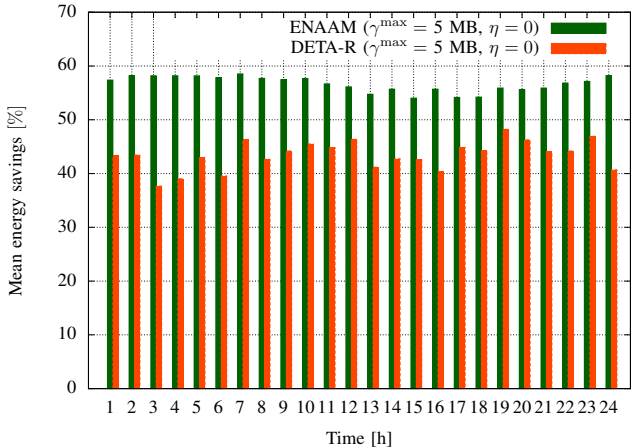
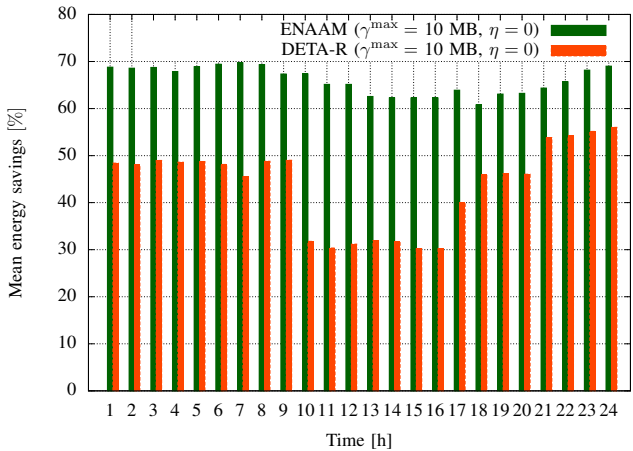
\begin{figure}[t]
	\centering
	\begin{subfigure}[t]{\columnwidth}
		\centering
		\resizebox{\columnwidth}{!}{\input{savings_06.tex}}
		\caption{Mean energy savings for $\eta = 0$ and $\gamma^{\max} = 5$ MB.}
		\label{fig:savings_05}	
	\end{subfigure}
	\quad
	\begin{subfigure}[t]{\columnwidth}
		\centering
		\resizebox{\columnwidth}{!}{\input{savings_10.tex}}
		\caption{Mean energy savings for $\eta = 0$ and $\gamma^{\max} = 10$ MB.}
		\label{fig:savings_010}	
	\end{subfigure}
	\centering
	\caption{Mean energy savings for the single BS case.}
	\label{figure:energy_sav_singleBS}
\end{figure}

\noindent \textbf{Single communication site}: Figs.~\ref{fig:savings_05} and~\ref{fig:savings_010} are computed with $\eta = 0$ using Cluster 1 and Solar 1 as traffic load and harvested energy profiles for each BS (see Figs.~\ref{fig:trace_load} and~\ref{fig:energy_trace}). Moreover, $\gamma^{\max} = 5$ MB and $10$ MB, respectively. They show the mean energy savings achieved over time when \mbox{on-demand} and \mbox{energy-aware} edge resource provisioning is enabled (i.e., BS sleep modes and VM \mbox{soft-scaling}), in comparison with the case where they are not applied. Our edge network management algorithm (\ac{ENAAM}) is benchmarked with another one that heuristically selects the amount of traffic that is to be processed locally, $B_n(t) \leq \Gamma_{n}(t)$, depending on the expected load behavior. It is named Dynamic and \mbox{Energy-Traffic-Aware} algorithm with Random behavior \mbox{({DETA-R})}. Both ENAAM and \mbox{DETA-R} are aware of the predictions in future time slots (see Section~\ref{predict}), however, \mbox{DETA-R} provisions edge resources using a heuristic scheme. \mbox{DETA-R} heuristic works as follows:  if the expected load difference is \mbox{$\hat{L}(t+1) - \hat{L}(t) >0$}, then the normalize workload to be processed by BS $n$ in the current time slot $t$, $B_n(t)$, is randomly selected in the range $[0.6, 1]$, otherwise, it is picked evenly at random in the range $(0,0.6)$. 

%

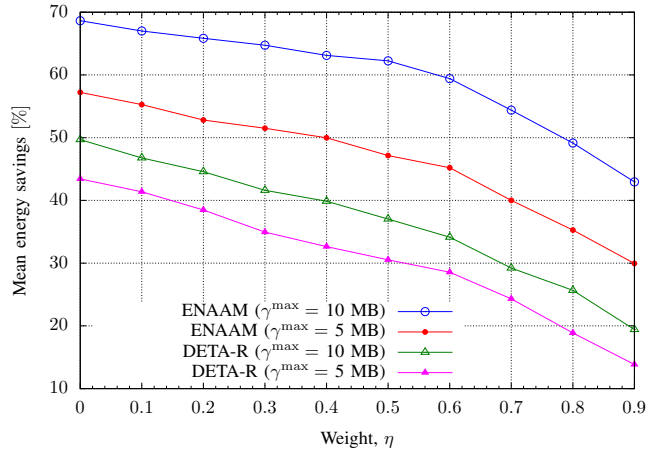
\begin{figure}[t]
	\centering
	\resizebox{\columnwidth}{!}{\input{single_bs.tex}}
	\caption{Energy savings {\it vs} weight $\eta$ (single BS case).}
	\label{fig:eta_val}
\end{figure} 

Average results for the ENAAM scheme show energy savings of $69\%$ ($\gamma^{\max} = 10$~MB) and $57\%$ ($\gamma^{\max} = 5$~MB), while \mbox{DETA-R} achieves $49\%$ ($\gamma^{\max} = 10$~MB) and $43\%$ ($\gamma^{\max} = 5$~MB) on average, where these savings are with respect to the case where {\it no energy management} is performed, i.e., the network is dimensioned for maximum expected capacity (maximum value of $\theta_{{\rm tot}, n}(t)$, with $M = 27$ \acp{VM}, $\forall \, t$). The results show that the maximum load allocated to each \ac{VM}, $\gamma^{\max}$, has an impact towards energy savings. An increase in energy savings is observed when $\gamma^{\max} = 10$~MB due to the fact that the number of \acp{VM} demanded per time slot is reduced, when compared to the allocation of $\gamma^{\max} = 5$~MB.

The \acp{ES} evolution with respect to $\eta$ is presented in Fig.~\ref{fig:eta_val}, taking into account the load allocated to each \ac{VM}, $\gamma^{\max}$. The results were obtained using Cluster 1 and Solar 1 as traffic load and harvested energy profiles ({\it see} Fig.~\ref{fig:trace_load} and Fig.~\ref{fig:energy_trace}). As expected, a drop in energy savings is observed when \ac{QoS} is prioritized, i.e., $\eta \to 1$, as in this case the BS energy consumption is no longer considered. It can be observed that ENAAM achieves a $50\%$ (or above) from $\eta = [0,0.4]$ when $\gamma^{\max} = 5$ MB and from  $\eta = [0,0.7]$ when $\gamma^{\max} = 10$ MB. This shows that the higher the load allocated to each \ac{VM}, the lesser the energy that is drained, as few \acp{VM} are running. \mbox{DETA-R} operates at below $50\%$ for all $\eta$ and $\gamma^{\max}$ values.\\

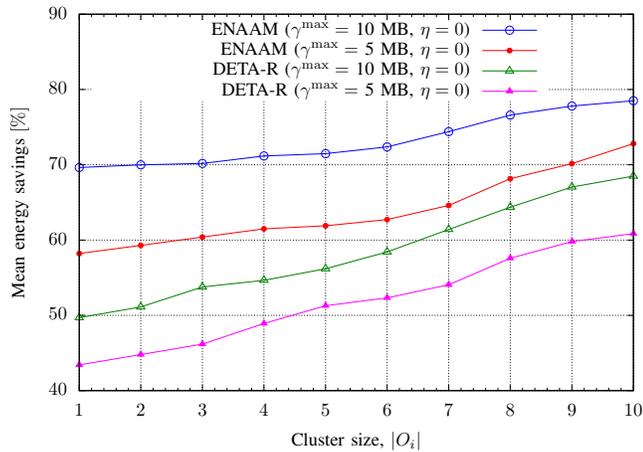
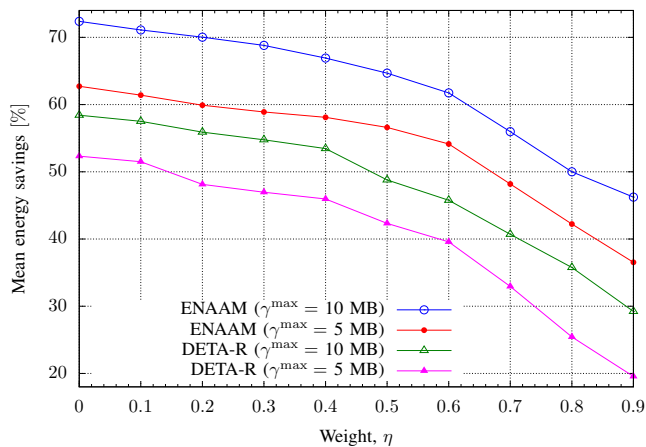
\begin{figure}[t]
	\centering
	\begin{subfigure}[t]{\columnwidth}
		\centering
		\resizebox{\columnwidth}{!}{\input{clusterplot_multibs.tex}}
		\caption{Energy savings {\it vs} cluster size.}
		\label{fig:cluster_size}	
	\end{subfigure}
	\quad
	\begin{subfigure}[t]{\columnwidth}
		\centering
		\resizebox{\columnwidth}{!}{\input{weight_plot.tex}}
		\caption{Energy savings {\it vs} $\eta$ for $|O_i|=6$.}
		\label{fig:cluster_weight}	
	\end{subfigure}
	\centering
	\caption{Energy savings for the multiple BSs case.}
	\label{figure:energy_sav_multBS}
\end{figure}

\noindent \textbf{Multiple communication sites}: Figs.~\ref{fig:cluster_size} and~\ref{fig:cluster_weight} present the mean energy savings achieved with respect to the cluster size and the weight $\eta$, using all the traffic load and harvested energy profiles from Figs.~\ref{fig:trace_load} and Fig.~\ref{fig:energy_trace}. Each BS randomly picks its own traffic load and harvested energy profile at the beginning of the optimization process. Here, to select the BS to be switched off, we use the management procedure of section~\ref{sub:site2_manager}. As for \mbox{DETA-R}, a BS is randomly selected to evolve its operating mode to power saving mode and offload its load to a nearby BS (in this case, the least loaded neighboring BS is selected), without taking into account its network impact measure.

Fig.~\ref{fig:cluster_size} shows the average energy savings obtained when clustering is adopted, i.e., here, the cluster size is increased from $|O_i| = 1$ to $10$ and $\eta = 0$. The obtained energy savings are with respect to the case where all BSs are dimensioned for maximum expected capacity (maximum value of $\theta_{{\rm tot}, n}(t)$, with $M$ = $27$ \acp{VM}, $\forall \, t, \forall \, n \in O_i$). It should be noted that the energy savings increase as the size of the cluster grows, thanks to the load balancing among active \acp{BS}, which cannot be implemented in the single communication site scenario (i.e., when BSs are independently managed). 

Then, Fig.~\ref{fig:cluster_weight} shows the average energy savings with respect to $\eta$, when the cluster size is set to an intermediate case ($| O_i|=6$). Again, here the energy savings are obtained with respect to the case where all the BSs are dimensioned for maximum capacity.
As expected, there is a drop in the energy savings achieved as the value of $\eta$ increases, as \ac{QoS} is prioritized. It can be observed that ENAAM achieves a value of $50\%$ or above when $\eta = [0,0.8]$ (at \mbox{$\gamma^{\max} = 10$~MB}) and when $\eta = [0,0.6]$ (at \mbox{$\gamma^{\max} = 5$~MB}). \mbox{DETA-R} achieves value above $50\%$ or above when $\eta = [0,0.4]$ (at $\gamma^{\max} = 10$) and $\eta = [0,0.1]$(at $\gamma^{\max} = 5$~MB).

Comparing Figs.~\ref{fig:eta_val} and~\ref{fig:cluster_weight}, an average gain of $9\%$ on the energy savings is observed when clustering is applied, by considering the mean energy savings with respect $\eta$ achieved with ENAAM for both cases. 	From Fig.~\ref{fig:cluster_size} we see that this gain can be as high as $16\%$ for ENAAM with $\gamma^{\max}=5$~MB (red curve) and bigger for the \mbox{DETA-R} approach. These results support the notion that performing a \mbox{clustering-based} optimization is beneficial thanks to the additional cooperation within each neighborhood of \acp{BS}. This cooperation allows to switch off more \acp{BS} through load balancing, increasing the energy savings while still controlling the users' \ac{QoS}.


\section{Conclusions}
\label{sec:concl}

In this paper, we have envisioned an edge network where a group of BSs are managed by a controller, for ease of BS organization and management, and also a mobile network where the edge apparatuses are powered by hybrid supplies, i.e., using green energy in order to promote energy \mbox{self-sustainability} and the power grid as a backup. Within the edge, each BS is endowed with computation capabilities to guarantee low latency to mobile users, offloading their workloads locally. The combination of energy saving methods, namely, \ac{BS} sleep modes and \ac{VM} \mbox{soft-scaling}, for single and multiple BS sites helps to reduce the mobile network's energy consumption. An edge energy management algorithm based on forecasting, clustering, control theory and heuristics, is proposed with the objective of saving energy within the access network, possibly making the BS system \mbox{self-sustainable}. Numerical results, obtained with \mbox{real-world} energy and traffic load traces, demonstrate that the proposed algorithm achieves energy savings between $57\%$ and $69\%$, on average, for the single communication site case, and a gain ranging from $9\%$ to $16\%$ on energy savings is observed when clustering is applied, with respect to the allocated maximum \mbox{per-VM} loads of $5$~MB and $10$~MB. The energy saving results are obtained with respect to the case where no energy management techniques are applied, either in one BS or single cluster.

\section*{Acknowledgements}

This work has received funding from the European Union's Horizon 2020 research and innovation programme under the Marie Sklodowska-Curie grant agreement No. 675891 \mbox{(SCAVENGE)}.

\bibliographystyle{IEEEtran}
\scriptsize
\bibliography{biblio}
\end{document}

%% file: traffic_profiles.tex
\begingroup
  \makeatletter
  \providecommand\color[2][]{%
    \GenericError{(gnuplot) \space\space\space\@spaces}{%
      Package color not loaded in conjunction with
      terminal option `colourtext'%
    }{See the gnuplot documentation for explanation.%
    }{Either use 'blacktext' in gnuplot or load the package
      color.sty in LaTeX.}%
    \renewcommand\color[2][]{}%
  }%
  \providecommand\includegraphics[2][]{%
    \GenericError{(gnuplot) \space\space\space\@spaces}{%
      Package graphicx or graphics not loaded%
    }{See the gnuplot documentation for explanation.%
    }{The gnuplot epslatex terminal needs graphicx.sty or graphics.sty.}%
    \renewcommand\includegraphics[2][]{}%
  }%
  \providecommand\rotatebox[2]{#2}%
  \@ifundefined{ifGPcolor}{%
    \newif\ifGPcolor
    \GPcolortrue
  }{}%
  \@ifundefined{ifGPblacktext}{%
    \newif\ifGPblacktext
    \GPblacktexttrue
  }{}%
  \let\gplgaddtomacro\g@addto@macro
  \gdef\gplbacktext{}%
  \gdef\gplfronttext{}%
  \makeatother
  \ifGPblacktext
    \def\colorrgb#1{}%
    \def\colorgray#1{}%
  \else
    \ifGPcolor
      \def\colorrgb#1{\color[rgb]{#1}}%
      \def\colorgray#1{\color[gray]{#1}}%
      \expandafter\def\csname LTw\endcsname{\color{white}}%
      \expandafter\def\csname LTb\endcsname{\color{black}}%
      \expandafter\def\csname LTa\endcsname{\color{black}}%
      \expandafter\def\csname LT0\endcsname{\color[rgb]{1,0,0}}%
      \expandafter\def\csname LT1\endcsname{\color[rgb]{0,1,0}}%
      \expandafter\def\csname LT2\endcsname{\color[rgb]{0,0,1}}%
      \expandafter\def\csname LT3\endcsname{\color[rgb]{1,0,1}}%
      \expandafter\def\csname LT4\endcsname{\color[rgb]{0,1,1}}%
      \expandafter\def\csname LT5\endcsname{\color[rgb]{1,1,0}}%
      \expandafter\def\csname LT6\endcsname{\color[rgb]{0,0,0}}%
      \expandafter\def\csname LT7\endcsname{\color[rgb]{1,0.3,0}}%
      \expandafter\def\csname LT8\endcsname{\color[rgb]{0.5,0.5,0.5}}%
    \else
      \def\colorrgb#1{\color{black}}%
      \def\colorgray#1{\color[gray]{#1}}%
      \expandafter\def\csname LTw\endcsname{\color{white}}%
      \expandafter\def\csname LTb\endcsname{\color{black}}%
      \expandafter\def\csname LTa\endcsname{\color{black}}%
      \expandafter\def\csname LT0\endcsname{\color{black}}%
      \expandafter\def\csname LT1\endcsname{\color{black}}%
      \expandafter\def\csname LT2\endcsname{\color{black}}%
      \expandafter\def\csname LT3\endcsname{\color{black}}%
      \expandafter\def\csname LT4\endcsname{\color{black}}%
      \expandafter\def\csname LT5\endcsname{\color{black}}%
      \expandafter\def\csname LT6\endcsname{\color{black}}%
      \expandafter\def\csname LT7\endcsname{\color{black}}%
      \expandafter\def\csname LT8\endcsname{\color{black}}%
    \fi
  \fi
    \setlength{\unitlength}{0.0500bp}%
    \ifx\gptboxheight\undefined%
      \newlength{\gptboxheight}%
      \newlength{\gptboxwidth}%
      \newsavebox{\gptboxtext}%
    \fi%
    \setlength{\fboxrule}{0.5pt}%
    \setlength{\fboxsep}{1pt}%
\begin{picture}(7200.00,5040.00)%
    \gplgaddtomacro\gplbacktext{%
      \csname LTb\endcsname%
      \put(946,704){\makebox(0,0)[r]{\strut{}$0$}}%
      \csname LTb\endcsname%
      \put(946,1518){\makebox(0,0)[r]{\strut{}$0.2$}}%
      \csname LTb\endcsname%
      \put(946,2332){\makebox(0,0)[r]{\strut{}$0.4$}}%
      \csname LTb\endcsname%
      \put(946,3147){\makebox(0,0)[r]{\strut{}$0.6$}}%
      \csname LTb\endcsname%
      \put(946,3961){\makebox(0,0)[r]{\strut{}$0.8$}}%
      \csname LTb\endcsname%
      \put(946,4775){\makebox(0,0)[r]{\strut{}$1$}}%
      \csname LTb\endcsname%
      \put(1565,484){\makebox(0,0){\strut{}$5$}}%
      \csname LTb\endcsname%
      \put(2174,484){\makebox(0,0){\strut{}$10$}}%
      \csname LTb\endcsname%
      \put(2783,484){\makebox(0,0){\strut{}$15$}}%
      \csname LTb\endcsname%
      \put(3392,484){\makebox(0,0){\strut{}$20$}}%
      \csname LTb\endcsname%
      \put(4001,484){\makebox(0,0){\strut{}$25$}}%
      \csname LTb\endcsname%
      \put(4610,484){\makebox(0,0){\strut{}$30$}}%
      \csname LTb\endcsname%
      \put(5219,484){\makebox(0,0){\strut{}$35$}}%
      \csname LTb\endcsname%
      \put(5829,484){\makebox(0,0){\strut{}$40$}}%
      \csname LTb\endcsname%
      \put(6438,484){\makebox(0,0){\strut{}$45$}}%
    }%
    \gplgaddtomacro\gplfronttext{%
      \csname LTb\endcsname%
      \put(176,2739){\rotatebox{-270}{\makebox(0,0){\strut{}Normalized traffic load}}}%
      \put(3940,154){\makebox(0,0){\strut{}Time [h]}}%
      \csname LTb\endcsname%
      \put(4974,1815){\makebox(0,0)[r]{\strut{}Cluster 1}}%
      \csname LTb\endcsname%
      \put(4974,1595){\makebox(0,0)[r]{\strut{}Cluster 2}}%
      \csname LTb\endcsname%
      \put(4974,1375){\makebox(0,0)[r]{\strut{}Cluster 3}}%
      \csname LTb\endcsname%
      \put(4974,1155){\makebox(0,0)[r]{\strut{}Cluster 4}}%
    }%
    \gplbacktext
    \put(0,0){\includegraphics{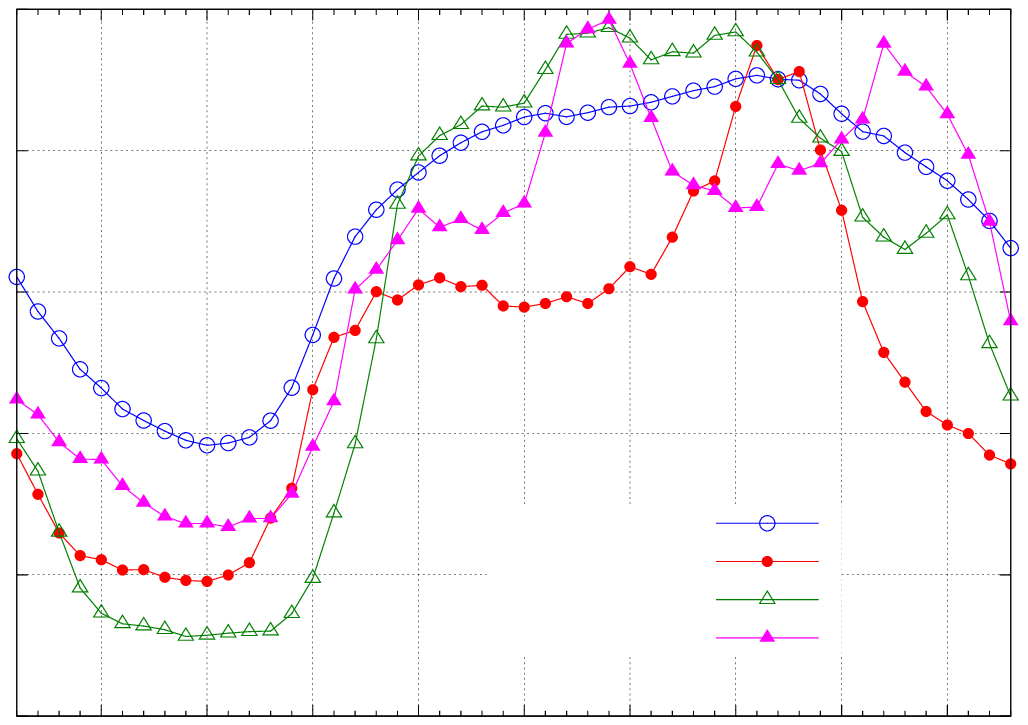}}%
    \gplfronttext
  \end{picture}%
\endgroup

%% file: energy_profiles.tex
\begingroup
  \makeatletter
  \providecommand\color[2][]{%
    \GenericError{(gnuplot) \space\space\space\@spaces}{%
      Package color not loaded in conjunction with
      terminal option `colourtext'%
    }{See the gnuplot documentation for explanation.%
    }{Either use 'blacktext' in gnuplot or load the package
      color.sty in LaTeX.}%
    \renewcommand\color[2][]{}%
  }%
  \providecommand\includegraphics[2][]{%
    \GenericError{(gnuplot) \space\space\space\@spaces}{%
      Package graphicx or graphics not loaded%
    }{See the gnuplot documentation for explanation.%
    }{The gnuplot epslatex terminal needs graphicx.sty or graphics.sty.}%
    \renewcommand\includegraphics[2][]{}%
  }%
  \providecommand\rotatebox[2]{#2}%
  \@ifundefined{ifGPcolor}{%
    \newif\ifGPcolor
    \GPcolortrue
  }{}%
  \@ifundefined{ifGPblacktext}{%
    \newif\ifGPblacktext
    \GPblacktexttrue
  }{}%
  \let\gplgaddtomacro\g@addto@macro
  \gdef\gplbacktext{}%
  \gdef\gplfronttext{}%
  \makeatother
  \ifGPblacktext
    \def\colorrgb#1{}%
    \def\colorgray#1{}%
  \else
    \ifGPcolor
      \def\colorrgb#1{\color[rgb]{#1}}%
      \def\colorgray#1{\color[gray]{#1}}%
      \expandafter\def\csname LTw\endcsname{\color{white}}%
      \expandafter\def\csname LTb\endcsname{\color{black}}%
      \expandafter\def\csname LTa\endcsname{\color{black}}%
      \expandafter\def\csname LT0\endcsname{\color[rgb]{1,0,0}}%
      \expandafter\def\csname LT1\endcsname{\color[rgb]{0,1,0}}%
      \expandafter\def\csname LT2\endcsname{\color[rgb]{0,0,1}}%
      \expandafter\def\csname LT3\endcsname{\color[rgb]{1,0,1}}%
      \expandafter\def\csname LT4\endcsname{\color[rgb]{0,1,1}}%
      \expandafter\def\csname LT5\endcsname{\color[rgb]{1,1,0}}%
      \expandafter\def\csname LT6\endcsname{\color[rgb]{0,0,0}}%
      \expandafter\def\csname LT7\endcsname{\color[rgb]{1,0.3,0}}%
      \expandafter\def\csname LT8\endcsname{\color[rgb]{0.5,0.5,0.5}}%
    \else
      \def\colorrgb#1{\color{black}}%
      \def\colorgray#1{\color[gray]{#1}}%
      \expandafter\def\csname LTw\endcsname{\color{white}}%
      \expandafter\def\csname LTb\endcsname{\color{black}}%
      \expandafter\def\csname LTa\endcsname{\color{black}}%
      \expandafter\def\csname LT0\endcsname{\color{black}}%
      \expandafter\def\csname LT1\endcsname{\color{black}}%
      \expandafter\def\csname LT2\endcsname{\color{black}}%
      \expandafter\def\csname LT3\endcsname{\color{black}}%
      \expandafter\def\csname LT4\endcsname{\color{black}}%
      \expandafter\def\csname LT5\endcsname{\color{black}}%
      \expandafter\def\csname LT6\endcsname{\color{black}}%
      \expandafter\def\csname LT7\endcsname{\color{black}}%
      \expandafter\def\csname LT8\endcsname{\color{black}}%
    \fi
  \fi
    \setlength{\unitlength}{0.0500bp}%
    \ifx\gptboxheight\undefined%
      \newlength{\gptboxheight}%
      \newlength{\gptboxwidth}%
      \newsavebox{\gptboxtext}%
    \fi%
    \setlength{\fboxrule}{0.5pt}%
    \setlength{\fboxsep}{1pt}%
\begin{picture}(7200.00,5040.00)%
    \gplgaddtomacro\gplbacktext{%
      \csname LTb\endcsname%
      \put(682,704){\makebox(0,0)[r]{\strut{}$0$}}%
      \csname LTb\endcsname%
      \put(682,1518){\makebox(0,0)[r]{\strut{}$0.2$}}%
      \csname LTb\endcsname%
      \put(682,2332){\makebox(0,0)[r]{\strut{}$0.4$}}%
      \csname LTb\endcsname%
      \put(682,3147){\makebox(0,0)[r]{\strut{}$0.6$}}%
      \csname LTb\endcsname%
      \put(682,3961){\makebox(0,0)[r]{\strut{}$0.8$}}%
      \csname LTb\endcsname%
      \put(682,4775){\makebox(0,0)[r]{\strut{}$1$}}%
      \csname LTb\endcsname%
      \put(814,484){\makebox(0,0){\strut{}1:00}}%
      \csname LTb\endcsname%
      \put(1844,484){\makebox(0,0){\strut{}5:00}}%
      \csname LTb\endcsname%
      \put(3132,484){\makebox(0,0){\strut{}10:00}}%
      \csname LTb\endcsname%
      \put(4420,484){\makebox(0,0){\strut{}15:00}}%
      \csname LTb\endcsname%
      \put(5708,484){\makebox(0,0){\strut{}20:00}}%
      \csname LTb\endcsname%
      \put(6738,484){\makebox(0,0){\strut{}24:00}}%
    }%
    \gplgaddtomacro\gplfronttext{%
      \csname LTb\endcsname%
      \put(176,2739){\rotatebox{-270}{\makebox(0,0){\strut{}Normalized harvested energy}}}%
      \put(3776,154){\makebox(0,0){\strut{}Time [h]}}%
      \csname LTb\endcsname%
      \put(5751,4602){\makebox(0,0)[r]{\strut{}Solar 1}}%
      \csname LTb\endcsname%
      \put(5751,4382){\makebox(0,0)[r]{\strut{}Solar 2}}%
      \csname LTb\endcsname%
      \put(5751,4162){\makebox(0,0)[r]{\strut{}Solar 3}}%
    }%
    \gplbacktext
    \put(0,0){\includegraphics{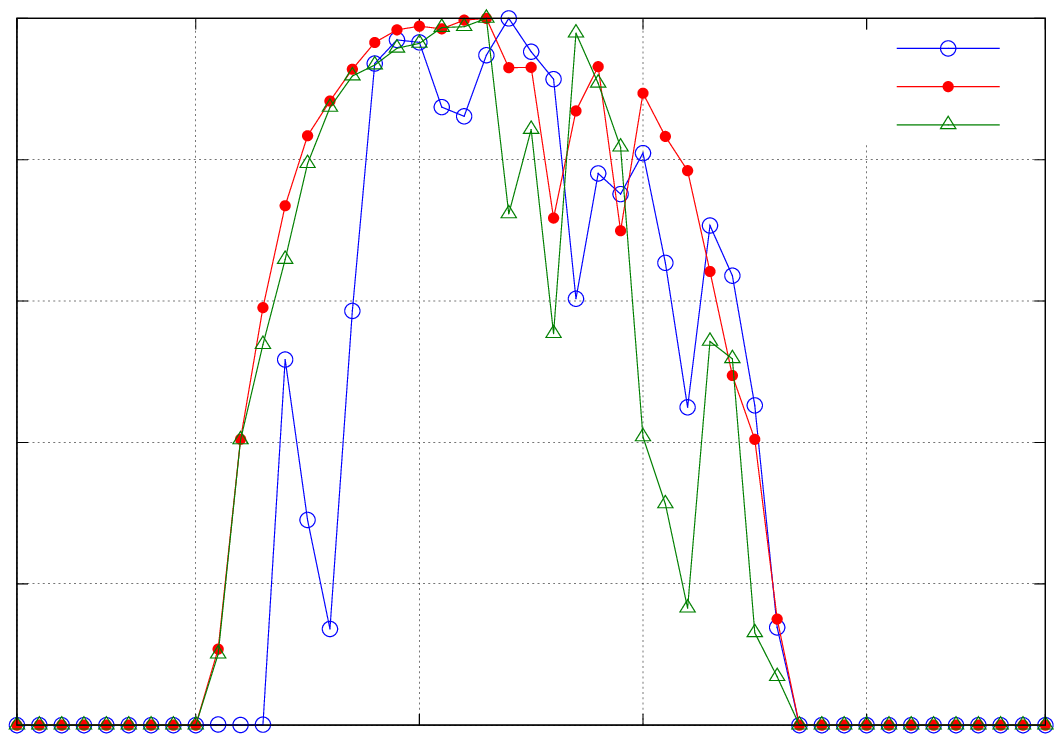}}%
    \gplfronttext
  \end{picture}%
\endgroup

%% file: load_pred_info.tex
\begingroup
  \makeatletter
  \providecommand\color[2][]{%
    \GenericError{(gnuplot) \space\space\space\@spaces}{%
      Package color not loaded in conjunction with
      terminal option `colourtext'%
    }{See the gnuplot documentation for explanation.%
    }{Either use 'blacktext' in gnuplot or load the package
      color.sty in LaTeX.}%
    \renewcommand\color[2][]{}%
  }%
  \providecommand\includegraphics[2][]{%
    \GenericError{(gnuplot) \space\space\space\@spaces}{%
      Package graphicx or graphics not loaded%
    }{See the gnuplot documentation for explanation.%
    }{The gnuplot epslatex terminal needs graphicx.sty or graphics.sty.}%
    \renewcommand\includegraphics[2][]{}%
  }%
  \providecommand\rotatebox[2]{#2}%
  \@ifundefined{ifGPcolor}{%
    \newif\ifGPcolor
    \GPcolortrue
  }{}%
  \@ifundefined{ifGPblacktext}{%
    \newif\ifGPblacktext
    \GPblacktexttrue
  }{}%
  \let\gplgaddtomacro\g@addto@macro
  \gdef\gplbacktext{}%
  \gdef\gplfronttext{}%
  \makeatother
  \ifGPblacktext
    \def\colorrgb#1{}%
    \def\colorgray#1{}%
  \else
    \ifGPcolor
      \def\colorrgb#1{\color[rgb]{#1}}%
      \def\colorgray#1{\color[gray]{#1}}%
      \expandafter\def\csname LTw\endcsname{\color{white}}%
      \expandafter\def\csname LTb\endcsname{\color{black}}%
      \expandafter\def\csname LTa\endcsname{\color{black}}%
      \expandafter\def\csname LT0\endcsname{\color[rgb]{1,0,0}}%
      \expandafter\def\csname LT1\endcsname{\color[rgb]{0,1,0}}%
      \expandafter\def\csname LT2\endcsname{\color[rgb]{0,0,1}}%
      \expandafter\def\csname LT3\endcsname{\color[rgb]{1,0,1}}%
      \expandafter\def\csname LT4\endcsname{\color[rgb]{0,1,1}}%
      \expandafter\def\csname LT5\endcsname{\color[rgb]{1,1,0}}%
      \expandafter\def\csname LT6\endcsname{\color[rgb]{0,0,0}}%
      \expandafter\def\csname LT7\endcsname{\color[rgb]{1,0.3,0}}%
      \expandafter\def\csname LT8\endcsname{\color[rgb]{0.5,0.5,0.5}}%
    \else
      \def\colorrgb#1{\color{black}}%
      \def\colorgray#1{\color[gray]{#1}}%
      \expandafter\def\csname LTw\endcsname{\color{white}}%
      \expandafter\def\csname LTb\endcsname{\color{black}}%
      \expandafter\def\csname LTa\endcsname{\color{black}}%
      \expandafter\def\csname LT0\endcsname{\color{black}}%
      \expandafter\def\csname LT1\endcsname{\color{black}}%
      \expandafter\def\csname LT2\endcsname{\color{black}}%
      \expandafter\def\csname LT3\endcsname{\color{black}}%
      \expandafter\def\csname LT4\endcsname{\color{black}}%
      \expandafter\def\csname LT5\endcsname{\color{black}}%
      \expandafter\def\csname LT6\endcsname{\color{black}}%
      \expandafter\def\csname LT7\endcsname{\color{black}}%
      \expandafter\def\csname LT8\endcsname{\color{black}}%
    \fi
  \fi
    \setlength{\unitlength}{0.0500bp}%
    \ifx\gptboxheight\undefined%
      \newlength{\gptboxheight}%
      \newlength{\gptboxwidth}%
      \newsavebox{\gptboxtext}%
    \fi%
    \setlength{\fboxrule}{0.5pt}%
    \setlength{\fboxsep}{1pt}%
\begin{picture}(7200.00,5040.00)%
    \gplgaddtomacro\gplbacktext{%
      \csname LTb\endcsname%
      \put(682,704){\makebox(0,0)[r]{\strut{}$0.3$}}%
      \csname LTb\endcsname%
      \put(682,1330){\makebox(0,0)[r]{\strut{}$0.4$}}%
      \csname LTb\endcsname%
      \put(682,1957){\makebox(0,0)[r]{\strut{}$0.5$}}%
      \csname LTb\endcsname%
      \put(682,2583){\makebox(0,0)[r]{\strut{}$0.6$}}%
      \csname LTb\endcsname%
      \put(682,3209){\makebox(0,0)[r]{\strut{}$0.7$}}%
      \csname LTb\endcsname%
      \put(682,3836){\makebox(0,0)[r]{\strut{}$0.8$}}%
      \csname LTb\endcsname%
      \put(682,4462){\makebox(0,0)[r]{\strut{}$0.9$}}%
      \csname LTb\endcsname%
      \put(941,484){\makebox(0,0){\strut{}1:00}}%
      \csname LTb\endcsname%
      \put(1961,484){\makebox(0,0){\strut{}5:00}}%
      \csname LTb\endcsname%
      \put(3235,484){\makebox(0,0){\strut{}10:00}}%
      \csname LTb\endcsname%
      \put(4509,484){\makebox(0,0){\strut{}15:00}}%
      \csname LTb\endcsname%
      \put(5784,484){\makebox(0,0){\strut{}20:00}}%
      \csname LTb\endcsname%
      \put(6803,484){\makebox(0,0){\strut{}24:00}}%
    }%
    \gplgaddtomacro\gplfronttext{%
      \csname LTb\endcsname%
      \put(176,2739){\rotatebox{-270}{\makebox(0,0){\strut{}Normalized traffic load}}}%
      \put(3808,154){\makebox(0,0){\strut{}Time [h]}}%
      \csname LTb\endcsname%
      \put(5816,1097){\makebox(0,0)[r]{\strut{}Real values}}%
      \csname LTb\endcsname%
      \put(5816,877){\makebox(0,0)[r]{\strut{}Predictions}}%
    }%
    \gplbacktext
    \put(0,0){\includegraphics{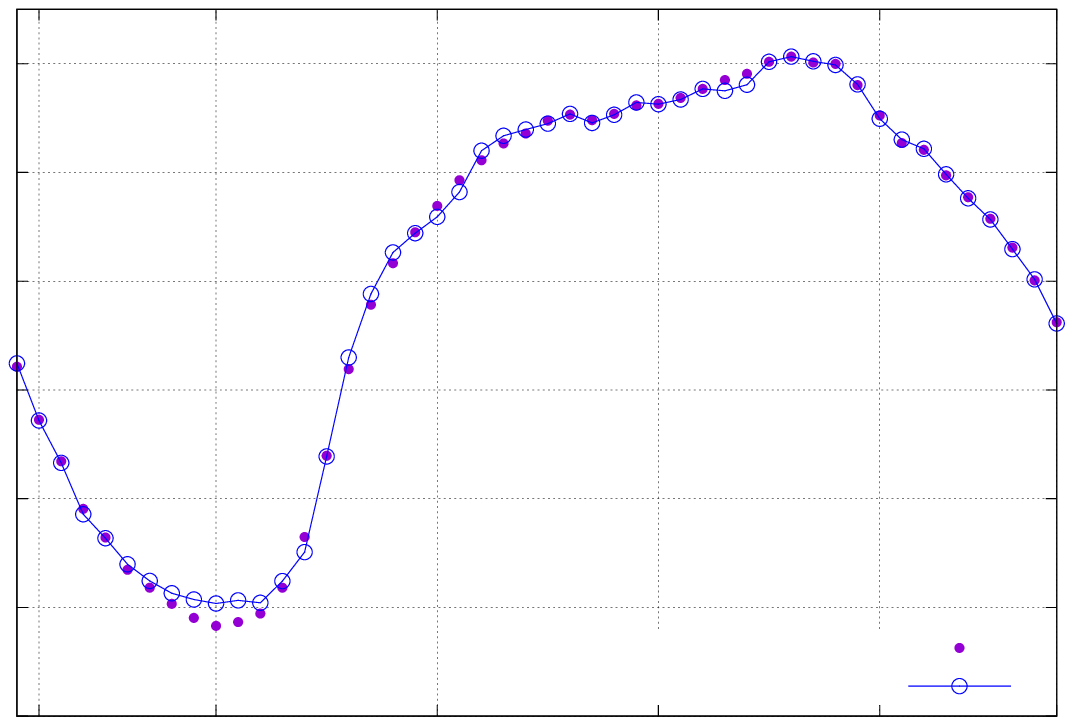}}%
    \gplfronttext
  \end{picture}%
\endgroup

%% file: energy_pred_info.tex
\begingroup
  \makeatletter
  \providecommand\color[2][]{%
    \GenericError{(gnuplot) \space\space\space\@spaces}{%
      Package color not loaded in conjunction with
      terminal option `colourtext'%
    }{See the gnuplot documentation for explanation.%
    }{Either use 'blacktext' in gnuplot or load the package
      color.sty in LaTeX.}%
    \renewcommand\color[2][]{}%
  }%
  \providecommand\includegraphics[2][]{%
    \GenericError{(gnuplot) \space\space\space\@spaces}{%
      Package graphicx or graphics not loaded%
    }{See the gnuplot documentation for explanation.%
    }{The gnuplot epslatex terminal needs graphicx.sty or graphics.sty.}%
    \renewcommand\includegraphics[2][]{}%
  }%
  \providecommand\rotatebox[2]{#2}%
  \@ifundefined{ifGPcolor}{%
    \newif\ifGPcolor
    \GPcolortrue
  }{}%
  \@ifundefined{ifGPblacktext}{%
    \newif\ifGPblacktext
    \GPblacktexttrue
  }{}%
  \let\gplgaddtomacro\g@addto@macro
  \gdef\gplbacktext{}%
  \gdef\gplfronttext{}%
  \makeatother
  \ifGPblacktext
    \def\colorrgb#1{}%
    \def\colorgray#1{}%
  \else
    \ifGPcolor
      \def\colorrgb#1{\color[rgb]{#1}}%
      \def\colorgray#1{\color[gray]{#1}}%
      \expandafter\def\csname LTw\endcsname{\color{white}}%
      \expandafter\def\csname LTb\endcsname{\color{black}}%
      \expandafter\def\csname LTa\endcsname{\color{black}}%
      \expandafter\def\csname LT0\endcsname{\color[rgb]{1,0,0}}%
      \expandafter\def\csname LT1\endcsname{\color[rgb]{0,1,0}}%
      \expandafter\def\csname LT2\endcsname{\color[rgb]{0,0,1}}%
      \expandafter\def\csname LT3\endcsname{\color[rgb]{1,0,1}}%
      \expandafter\def\csname LT4\endcsname{\color[rgb]{0,1,1}}%
      \expandafter\def\csname LT5\endcsname{\color[rgb]{1,1,0}}%
      \expandafter\def\csname LT6\endcsname{\color[rgb]{0,0,0}}%
      \expandafter\def\csname LT7\endcsname{\color[rgb]{1,0.3,0}}%
      \expandafter\def\csname LT8\endcsname{\color[rgb]{0.5,0.5,0.5}}%
    \else
      \def\colorrgb#1{\color{black}}%
      \def\colorgray#1{\color[gray]{#1}}%
      \expandafter\def\csname LTw\endcsname{\color{white}}%
      \expandafter\def\csname LTb\endcsname{\color{black}}%
      \expandafter\def\csname LTa\endcsname{\color{black}}%
      \expandafter\def\csname LT0\endcsname{\color{black}}%
      \expandafter\def\csname LT1\endcsname{\color{black}}%
      \expandafter\def\csname LT2\endcsname{\color{black}}%
      \expandafter\def\csname LT3\endcsname{\color{black}}%
      \expandafter\def\csname LT4\endcsname{\color{black}}%
      \expandafter\def\csname LT5\endcsname{\color{black}}%
      \expandafter\def\csname LT6\endcsname{\color{black}}%
      \expandafter\def\csname LT7\endcsname{\color{black}}%
      \expandafter\def\csname LT8\endcsname{\color{black}}%
    \fi
  \fi
    \setlength{\unitlength}{0.0500bp}%
    \ifx\gptboxheight\undefined%
      \newlength{\gptboxheight}%
      \newlength{\gptboxwidth}%
      \newsavebox{\gptboxtext}%
    \fi%
    \setlength{\fboxrule}{0.5pt}%
    \setlength{\fboxsep}{1pt}%
\begin{picture}(7200.00,5040.00)%
    \gplgaddtomacro\gplbacktext{%
      \csname LTb\endcsname%
      \put(682,704){\makebox(0,0)[r]{\strut{}$0$}}%
      \csname LTb\endcsname%
      \put(682,1518){\makebox(0,0)[r]{\strut{}$0.2$}}%
      \csname LTb\endcsname%
      \put(682,2332){\makebox(0,0)[r]{\strut{}$0.4$}}%
      \csname LTb\endcsname%
      \put(682,3147){\makebox(0,0)[r]{\strut{}$0.6$}}%
      \csname LTb\endcsname%
      \put(682,3961){\makebox(0,0)[r]{\strut{}$0.8$}}%
      \csname LTb\endcsname%
      \put(682,4775){\makebox(0,0)[r]{\strut{}$1$}}%
      \csname LTb\endcsname%
      \put(941,484){\makebox(0,0){\strut{}1:00}}%
      \csname LTb\endcsname%
      \put(1961,484){\makebox(0,0){\strut{}5:00}}%
      \csname LTb\endcsname%
      \put(3235,484){\makebox(0,0){\strut{}10:00}}%
      \csname LTb\endcsname%
      \put(4509,484){\makebox(0,0){\strut{}15:00}}%
      \csname LTb\endcsname%
      \put(5784,484){\makebox(0,0){\strut{}20:00}}%
      \csname LTb\endcsname%
      \put(6803,484){\makebox(0,0){\strut{}24:00}}%
    }%
    \gplgaddtomacro\gplfronttext{%
      \csname LTb\endcsname%
      \put(176,2739){\rotatebox{-270}{\makebox(0,0){\strut{}Normalized harvested energy}}}%
      \put(3808,154){\makebox(0,0){\strut{}Time [h]}}%
      \csname LTb\endcsname%
      \put(5816,4602){\makebox(0,0)[r]{\strut{}Real values}}%
      \csname LTb\endcsname%
      \put(5816,4382){\makebox(0,0)[r]{\strut{}Predictions}}%
    }%
    \gplbacktext
    \put(0,0){\includegraphics{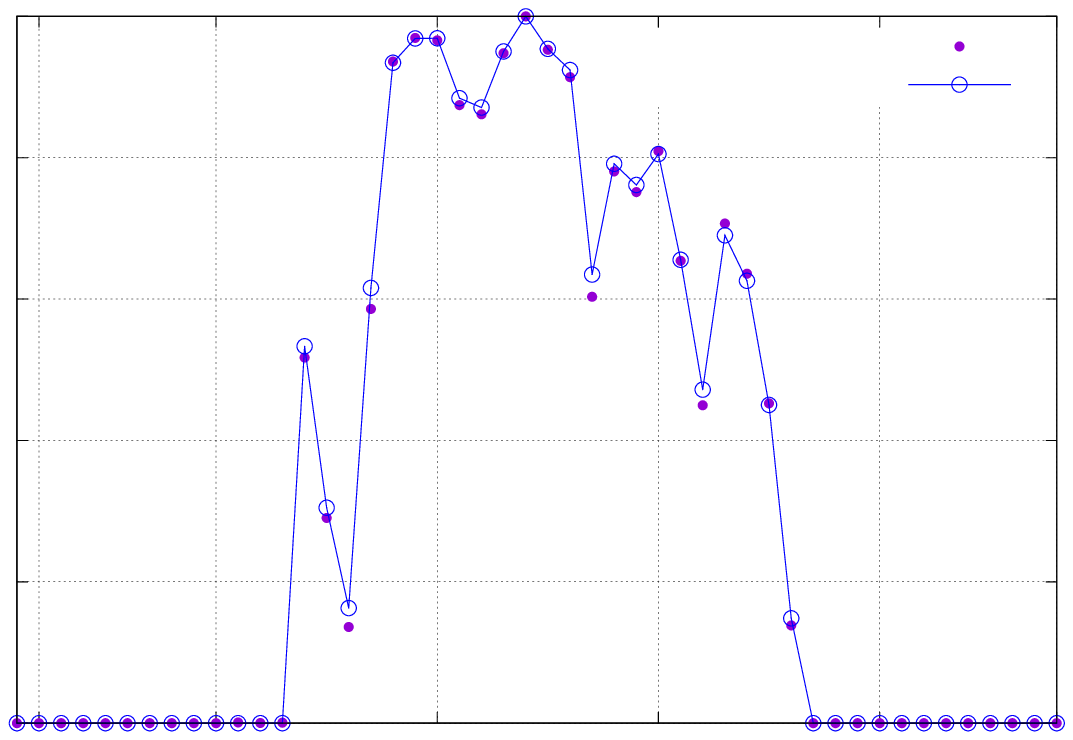}}%
    \gplfronttext
  \end{picture}%
\endgroup

%% file: savings_06.tex
\begingroup
  \makeatletter
  \providecommand\color[2][]{%
    \GenericError{(gnuplot) \space\space\space\@spaces}{%
      Package color not loaded in conjunction with
      terminal option `colourtext'%
    }{See the gnuplot documentation for explanation.%
    }{Either use 'blacktext' in gnuplot or load the package
      color.sty in LaTeX.}%
    \renewcommand\color[2][]{}%
  }%
  \providecommand\includegraphics[2][]{%
    \GenericError{(gnuplot) \space\space\space\@spaces}{%
      Package graphicx or graphics not loaded%
    }{See the gnuplot documentation for explanation.%
    }{The gnuplot epslatex terminal needs graphicx.sty or graphics.sty.}%
    \renewcommand\includegraphics[2][]{}%
  }%
  \providecommand\rotatebox[2]{#2}%
  \@ifundefined{ifGPcolor}{%
    \newif\ifGPcolor
    \GPcolortrue
  }{}%
  \@ifundefined{ifGPblacktext}{%
    \newif\ifGPblacktext
    \GPblacktexttrue
  }{}%
  \let\gplgaddtomacro\g@addto@macro
  \gdef\gplbacktext{}%
  \gdef\gplfronttext{}%
  \makeatother
  \ifGPblacktext
    \def\colorrgb#1{}%
    \def\colorgray#1{}%
  \else
    \ifGPcolor
      \def\colorrgb#1{\color[rgb]{#1}}%
      \def\colorgray#1{\color[gray]{#1}}%
      \expandafter\def\csname LTw\endcsname{\color{white}}%
      \expandafter\def\csname LTb\endcsname{\color{black}}%
      \expandafter\def\csname LTa\endcsname{\color{black}}%
      \expandafter\def\csname LT0\endcsname{\color[rgb]{1,0,0}}%
      \expandafter\def\csname LT1\endcsname{\color[rgb]{0,1,0}}%
      \expandafter\def\csname LT2\endcsname{\color[rgb]{0,0,1}}%
      \expandafter\def\csname LT3\endcsname{\color[rgb]{1,0,1}}%
      \expandafter\def\csname LT4\endcsname{\color[rgb]{0,1,1}}%
      \expandafter\def\csname LT5\endcsname{\color[rgb]{1,1,0}}%
      \expandafter\def\csname LT6\endcsname{\color[rgb]{0,0,0}}%
      \expandafter\def\csname LT7\endcsname{\color[rgb]{1,0.3,0}}%
      \expandafter\def\csname LT8\endcsname{\color[rgb]{0.5,0.5,0.5}}%
    \else
      \def\colorrgb#1{\color{black}}%
      \def\colorgray#1{\color[gray]{#1}}%
      \expandafter\def\csname LTw\endcsname{\color{white}}%
      \expandafter\def\csname LTb\endcsname{\color{black}}%
      \expandafter\def\csname LTa\endcsname{\color{black}}%
      \expandafter\def\csname LT0\endcsname{\color{black}}%
      \expandafter\def\csname LT1\endcsname{\color{black}}%
      \expandafter\def\csname LT2\endcsname{\color{black}}%
      \expandafter\def\csname LT3\endcsname{\color{black}}%
      \expandafter\def\csname LT4\endcsname{\color{black}}%
      \expandafter\def\csname LT5\endcsname{\color{black}}%
      \expandafter\def\csname LT6\endcsname{\color{black}}%
      \expandafter\def\csname LT7\endcsname{\color{black}}%
      \expandafter\def\csname LT8\endcsname{\color{black}}%
    \fi
  \fi
  \setlength{\unitlength}{0.0500bp}%
  \begin{picture}(7200.00,5040.00)%
    \gplgaddtomacro\gplbacktext{%
      \csname LTb\endcsname%
      \put(814,704){\makebox(0,0)[r]{\strut{} 0}}%
      \csname LTb\endcsname%
      \put(814,1286){\makebox(0,0)[r]{\strut{} 10}}%
      \csname LTb\endcsname%
      \put(814,1867){\makebox(0,0)[r]{\strut{} 20}}%
      \csname LTb\endcsname%
      \put(814,2449){\makebox(0,0)[r]{\strut{} 30}}%
      \csname LTb\endcsname%
      \put(814,3030){\makebox(0,0)[r]{\strut{} 40}}%
      \csname LTb\endcsname%
      \put(814,3612){\makebox(0,0)[r]{\strut{} 50}}%
      \csname LTb\endcsname%
      \put(814,4193){\makebox(0,0)[r]{\strut{} 60}}%
      \csname LTb\endcsname%
      \put(814,4775){\makebox(0,0)[r]{\strut{} 70}}%
      \csname LTb\endcsname%
      \put(1180,484){\makebox(0,0){\strut{}1}}%
      \csname LTb\endcsname%
      \put(1415,484){\makebox(0,0){\strut{}2}}%
      \csname LTb\endcsname%
      \put(1649,484){\makebox(0,0){\strut{}3}}%
      \csname LTb\endcsname%
      \put(1883,484){\makebox(0,0){\strut{}4}}%
      \csname LTb\endcsname%
      \put(2117,484){\makebox(0,0){\strut{}5}}%
      \csname LTb\endcsname%
      \put(2352,484){\makebox(0,0){\strut{}6}}%
      \csname LTb\endcsname%
      \put(2586,484){\makebox(0,0){\strut{}7}}%
      \csname LTb\endcsname%
      \put(2820,484){\makebox(0,0){\strut{}8}}%
      \csname LTb\endcsname%
      \put(3055,484){\makebox(0,0){\strut{}9}}%
      \csname LTb\endcsname%
      \put(3289,484){\makebox(0,0){\strut{}10}}%
      \csname LTb\endcsname%
      \put(3523,484){\makebox(0,0){\strut{}11}}%
      \csname LTb\endcsname%
      \put(3757,484){\makebox(0,0){\strut{}12}}%
      \csname LTb\endcsname%
      \put(3992,484){\makebox(0,0){\strut{}13}}%
      \csname LTb\endcsname%
      \put(4226,484){\makebox(0,0){\strut{}14}}%
      \csname LTb\endcsname%
      \put(4460,484){\makebox(0,0){\strut{}15}}%
      \csname LTb\endcsname%
      \put(4694,484){\makebox(0,0){\strut{}16}}%
      \csname LTb\endcsname%
      \put(4929,484){\makebox(0,0){\strut{}17}}%
      \csname LTb\endcsname%
      \put(5163,484){\makebox(0,0){\strut{}18}}%
      \csname LTb\endcsname%
      \put(5397,484){\makebox(0,0){\strut{}19}}%
      \csname LTb\endcsname%
      \put(5632,484){\makebox(0,0){\strut{}20}}%
      \csname LTb\endcsname%
      \put(5866,484){\makebox(0,0){\strut{}21}}%
      \csname LTb\endcsname%
      \put(6100,484){\makebox(0,0){\strut{}22}}%
      \csname LTb\endcsname%
      \put(6334,484){\makebox(0,0){\strut{}23}}%
      \csname LTb\endcsname%
      \put(6569,484){\makebox(0,0){\strut{}24}}%
      \put(176,2739){\rotatebox{-270}{\makebox(0,0){\strut{}Mean energy savings $[\%]$}}}%
      \put(3874,154){\makebox(0,0){\strut{}Time [h]}}%
    }%
    \gplgaddtomacro\gplfronttext{%
      \csname LTb\endcsname%
      \put(5816,4602){\makebox(0,0)[r]{\strut{}ENAAM ($\gamma^{\rm max} = $ 5 MB, $\eta = 0 $)}}%
      \csname LTb\endcsname%
      \put(5816,4382){\makebox(0,0)[r]{\strut{}DETA-R ($\gamma^{\rm max} = $ 5 MB, $\eta = 0 $)}}%
    }%
    \gplbacktext
    \put(0,0){\includegraphics{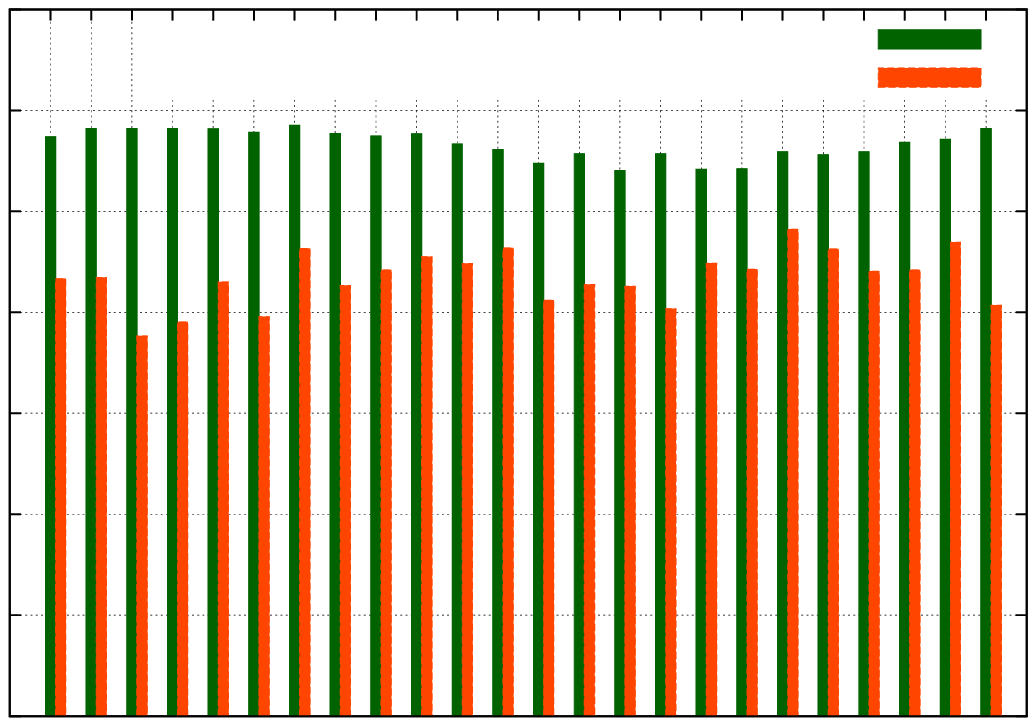}}%
    \gplfronttext
  \end{picture}%
\endgroup

%% file: savings_10.tex
\begingroup
  \makeatletter
  \providecommand\color[2][]{%
    \GenericError{(gnuplot) \space\space\space\@spaces}{%
      Package color not loaded in conjunction with
      terminal option `colourtext'%
    }{See the gnuplot documentation for explanation.%
    }{Either use 'blacktext' in gnuplot or load the package
      color.sty in LaTeX.}%
    \renewcommand\color[2][]{}%
  }%
  \providecommand\includegraphics[2][]{%
    \GenericError{(gnuplot) \space\space\space\@spaces}{%
      Package graphicx or graphics not loaded%
    }{See the gnuplot documentation for explanation.%
    }{The gnuplot epslatex terminal needs graphicx.sty or graphics.sty.}%
    \renewcommand\includegraphics[2][]{}%
  }%
  \providecommand\rotatebox[2]{#2}%
  \@ifundefined{ifGPcolor}{%
    \newif\ifGPcolor
    \GPcolortrue
  }{}%
  \@ifundefined{ifGPblacktext}{%
    \newif\ifGPblacktext
    \GPblacktexttrue
  }{}%
  \let\gplgaddtomacro\g@addto@macro
  \gdef\gplbacktext{}%
  \gdef\gplfronttext{}%
  \makeatother
  \ifGPblacktext
    \def\colorrgb#1{}%
    \def\colorgray#1{}%
  \else
    \ifGPcolor
      \def\colorrgb#1{\color[rgb]{#1}}%
      \def\colorgray#1{\color[gray]{#1}}%
      \expandafter\def\csname LTw\endcsname{\color{white}}%
      \expandafter\def\csname LTb\endcsname{\color{black}}%
      \expandafter\def\csname LTa\endcsname{\color{black}}%
      \expandafter\def\csname LT0\endcsname{\color[rgb]{1,0,0}}%
      \expandafter\def\csname LT1\endcsname{\color[rgb]{0,1,0}}%
      \expandafter\def\csname LT2\endcsname{\color[rgb]{0,0,1}}%
      \expandafter\def\csname LT3\endcsname{\color[rgb]{1,0,1}}%
      \expandafter\def\csname LT4\endcsname{\color[rgb]{0,1,1}}%
      \expandafter\def\csname LT5\endcsname{\color[rgb]{1,1,0}}%
      \expandafter\def\csname LT6\endcsname{\color[rgb]{0,0,0}}%
      \expandafter\def\csname LT7\endcsname{\color[rgb]{1,0.3,0}}%
      \expandafter\def\csname LT8\endcsname{\color[rgb]{0.5,0.5,0.5}}%
    \else
      \def\colorrgb#1{\color{black}}%
      \def\colorgray#1{\color[gray]{#1}}%
      \expandafter\def\csname LTw\endcsname{\color{white}}%
      \expandafter\def\csname LTb\endcsname{\color{black}}%
      \expandafter\def\csname LTa\endcsname{\color{black}}%
      \expandafter\def\csname LT0\endcsname{\color{black}}%
      \expandafter\def\csname LT1\endcsname{\color{black}}%
      \expandafter\def\csname LT2\endcsname{\color{black}}%
      \expandafter\def\csname LT3\endcsname{\color{black}}%
      \expandafter\def\csname LT4\endcsname{\color{black}}%
      \expandafter\def\csname LT5\endcsname{\color{black}}%
      \expandafter\def\csname LT6\endcsname{\color{black}}%
      \expandafter\def\csname LT7\endcsname{\color{black}}%
      \expandafter\def\csname LT8\endcsname{\color{black}}%
    \fi
  \fi
  \setlength{\unitlength}{0.0500bp}%
  \begin{picture}(7200.00,5040.00)%
    \gplgaddtomacro\gplbacktext{%
      \csname LTb\endcsname%
      \put(814,704){\makebox(0,0)[r]{\strut{} 0}}%
      \csname LTb\endcsname%
      \put(814,1213){\makebox(0,0)[r]{\strut{} 10}}%
      \csname LTb\endcsname%
      \put(814,1722){\makebox(0,0)[r]{\strut{} 20}}%
      \csname LTb\endcsname%
      \put(814,2231){\makebox(0,0)[r]{\strut{} 30}}%
      \csname LTb\endcsname%
      \put(814,2740){\makebox(0,0)[r]{\strut{} 40}}%
      \csname LTb\endcsname%
      \put(814,3248){\makebox(0,0)[r]{\strut{} 50}}%
      \csname LTb\endcsname%
      \put(814,3757){\makebox(0,0)[r]{\strut{} 60}}%
      \csname LTb\endcsname%
      \put(814,4266){\makebox(0,0)[r]{\strut{} 70}}%
      \csname LTb\endcsname%
      \put(814,4775){\makebox(0,0)[r]{\strut{} 80}}%
      \csname LTb\endcsname%
      \put(1180,484){\makebox(0,0){\strut{}1}}%
      \csname LTb\endcsname%
      \put(1415,484){\makebox(0,0){\strut{}2}}%
      \csname LTb\endcsname%
      \put(1649,484){\makebox(0,0){\strut{}3}}%
      \csname LTb\endcsname%
      \put(1883,484){\makebox(0,0){\strut{}4}}%
      \csname LTb\endcsname%
      \put(2117,484){\makebox(0,0){\strut{}5}}%
      \csname LTb\endcsname%
      \put(2352,484){\makebox(0,0){\strut{}6}}%
      \csname LTb\endcsname%
      \put(2586,484){\makebox(0,0){\strut{}7}}%
      \csname LTb\endcsname%
      \put(2820,484){\makebox(0,0){\strut{}8}}%
      \csname LTb\endcsname%
      \put(3055,484){\makebox(0,0){\strut{}9}}%
      \csname LTb\endcsname%
      \put(3289,484){\makebox(0,0){\strut{}10}}%
      \csname LTb\endcsname%
      \put(3523,484){\makebox(0,0){\strut{}11}}%
      \csname LTb\endcsname%
      \put(3757,484){\makebox(0,0){\strut{}12}}%
      \csname LTb\endcsname%
      \put(3992,484){\makebox(0,0){\strut{}13}}%
      \csname LTb\endcsname%
      \put(4226,484){\makebox(0,0){\strut{}14}}%
      \csname LTb\endcsname%
      \put(4460,484){\makebox(0,0){\strut{}15}}%
      \csname LTb\endcsname%
      \put(4694,484){\makebox(0,0){\strut{}16}}%
      \csname LTb\endcsname%
      \put(4929,484){\makebox(0,0){\strut{}17}}%
      \csname LTb\endcsname%
      \put(5163,484){\makebox(0,0){\strut{}18}}%
      \csname LTb\endcsname%
      \put(5397,484){\makebox(0,0){\strut{}19}}%
      \csname LTb\endcsname%
      \put(5632,484){\makebox(0,0){\strut{}20}}%
      \csname LTb\endcsname%
      \put(5866,484){\makebox(0,0){\strut{}21}}%
      \csname LTb\endcsname%
      \put(6100,484){\makebox(0,0){\strut{}22}}%
      \csname LTb\endcsname%
      \put(6334,484){\makebox(0,0){\strut{}23}}%
      \csname LTb\endcsname%
      \put(6569,484){\makebox(0,0){\strut{}24}}%
      \put(176,2739){\rotatebox{-270}{\makebox(0,0){\strut{}Mean energy savings $[\%]$}}}%
      \put(3874,154){\makebox(0,0){\strut{}Time [h]}}%
    }%
    \gplgaddtomacro\gplfronttext{%
      \csname LTb\endcsname%
      \put(5816,4602){\makebox(0,0)[r]{\strut{}ENAAM ($\gamma^{\rm max} = $ 10 MB, $\eta = 0 $)}}%
      \csname LTb\endcsname%
      \put(5816,4382){\makebox(0,0)[r]{\strut{}DETA-R ($\gamma^{\rm max} = $ 10 MB, $\eta = 0 $)}}%
    }%
    \gplbacktext
    \put(0,0){\includegraphics{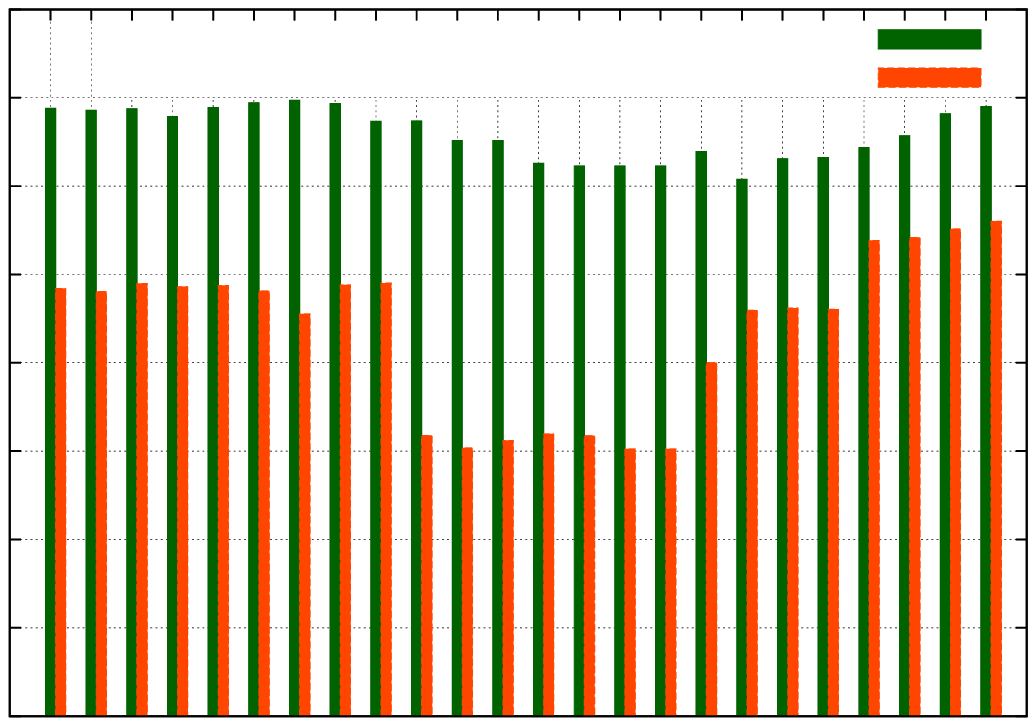}}%
    \gplfronttext
  \end{picture}%
\endgroup

%% file: single_bs.tex
\begingroup
  \makeatletter
  \providecommand\color[2][]{%
    \GenericError{(gnuplot) \space\space\space\@spaces}{%
      Package color not loaded in conjunction with
      terminal option `colourtext'%
    }{See the gnuplot documentation for explanation.%
    }{Either use 'blacktext' in gnuplot or load the package
      color.sty in LaTeX.}%
    \renewcommand\color[2][]{}%
  }%
  \providecommand\includegraphics[2][]{%
    \GenericError{(gnuplot) \space\space\space\@spaces}{%
      Package graphicx or graphics not loaded%
    }{See the gnuplot documentation for explanation.%
    }{The gnuplot epslatex terminal needs graphicx.sty or graphics.sty.}%
    \renewcommand\includegraphics[2][]{}%
  }%
  \providecommand\rotatebox[2]{#2}%
  \@ifundefined{ifGPcolor}{%
    \newif\ifGPcolor
    \GPcolortrue
  }{}%
  \@ifundefined{ifGPblacktext}{%
    \newif\ifGPblacktext
    \GPblacktexttrue
  }{}%
  \let\gplgaddtomacro\g@addto@macro
  \gdef\gplbacktext{}%
  \gdef\gplfronttext{}%
  \makeatother
  \ifGPblacktext
    \def\colorrgb#1{}%
    \def\colorgray#1{}%
  \else
    \ifGPcolor
      \def\colorrgb#1{\color[rgb]{#1}}%
      \def\colorgray#1{\color[gray]{#1}}%
      \expandafter\def\csname LTw\endcsname{\color{white}}%
      \expandafter\def\csname LTb\endcsname{\color{black}}%
      \expandafter\def\csname LTa\endcsname{\color{black}}%
      \expandafter\def\csname LT0\endcsname{\color[rgb]{1,0,0}}%
      \expandafter\def\csname LT1\endcsname{\color[rgb]{0,1,0}}%
      \expandafter\def\csname LT2\endcsname{\color[rgb]{0,0,1}}%
      \expandafter\def\csname LT3\endcsname{\color[rgb]{1,0,1}}%
      \expandafter\def\csname LT4\endcsname{\color[rgb]{0,1,1}}%
      \expandafter\def\csname LT5\endcsname{\color[rgb]{1,1,0}}%
      \expandafter\def\csname LT6\endcsname{\color[rgb]{0,0,0}}%
      \expandafter\def\csname LT7\endcsname{\color[rgb]{1,0.3,0}}%
      \expandafter\def\csname LT8\endcsname{\color[rgb]{0.5,0.5,0.5}}%
    \else
      \def\colorrgb#1{\color{black}}%
      \def\colorgray#1{\color[gray]{#1}}%
      \expandafter\def\csname LTw\endcsname{\color{white}}%
      \expandafter\def\csname LTb\endcsname{\color{black}}%
      \expandafter\def\csname LTa\endcsname{\color{black}}%
      \expandafter\def\csname LT0\endcsname{\color{black}}%
      \expandafter\def\csname LT1\endcsname{\color{black}}%
      \expandafter\def\csname LT2\endcsname{\color{black}}%
      \expandafter\def\csname LT3\endcsname{\color{black}}%
      \expandafter\def\csname LT4\endcsname{\color{black}}%
      \expandafter\def\csname LT5\endcsname{\color{black}}%
      \expandafter\def\csname LT6\endcsname{\color{black}}%
      \expandafter\def\csname LT7\endcsname{\color{black}}%
      \expandafter\def\csname LT8\endcsname{\color{black}}%
    \fi
  \fi
    \setlength{\unitlength}{0.0500bp}%
    \ifx\gptboxheight\undefined%
      \newlength{\gptboxheight}%
      \newlength{\gptboxwidth}%
      \newsavebox{\gptboxtext}%
    \fi%
    \setlength{\fboxrule}{0.5pt}%
    \setlength{\fboxsep}{1pt}%
\begin{picture}(7200.00,5040.00)%
    \gplgaddtomacro\gplbacktext{%
      \csname LTb\endcsname%
      \put(682,704){\makebox(0,0)[r]{\strut{}$10$}}%
      \csname LTb\endcsname%
      \put(682,1383){\makebox(0,0)[r]{\strut{}$20$}}%
      \csname LTb\endcsname%
      \put(682,2061){\makebox(0,0)[r]{\strut{}$30$}}%
      \csname LTb\endcsname%
      \put(682,2740){\makebox(0,0)[r]{\strut{}$40$}}%
      \csname LTb\endcsname%
      \put(682,3418){\makebox(0,0)[r]{\strut{}$50$}}%
      \csname LTb\endcsname%
      \put(682,4097){\makebox(0,0)[r]{\strut{}$60$}}%
      \csname LTb\endcsname%
      \put(682,4775){\makebox(0,0)[r]{\strut{}$70$}}%
      \csname LTb\endcsname%
      \put(814,484){\makebox(0,0){\strut{}$0$}}%
      \csname LTb\endcsname%
      \put(1479,484){\makebox(0,0){\strut{}$0.1$}}%
      \csname LTb\endcsname%
      \put(2145,484){\makebox(0,0){\strut{}$0.2$}}%
      \csname LTb\endcsname%
      \put(2810,484){\makebox(0,0){\strut{}$0.3$}}%
      \csname LTb\endcsname%
      \put(3476,484){\makebox(0,0){\strut{}$0.4$}}%
      \csname LTb\endcsname%
      \put(4141,484){\makebox(0,0){\strut{}$0.5$}}%
      \csname LTb\endcsname%
      \put(4807,484){\makebox(0,0){\strut{}$0.6$}}%
      \csname LTb\endcsname%
      \put(5472,484){\makebox(0,0){\strut{}$0.7$}}%
      \csname LTb\endcsname%
      \put(6138,484){\makebox(0,0){\strut{}$0.8$}}%
      \csname LTb\endcsname%
      \put(6803,484){\makebox(0,0){\strut{}$0.9$}}%
    }%
    \gplgaddtomacro\gplfronttext{%
      \csname LTb\endcsname%
      \put(176,2739){\rotatebox{-270}{\makebox(0,0){\strut{}Mean energy savings $[\%]$}}}%
      \put(3808,154){\makebox(0,0){\strut{}Weight, $\eta$}}%
      \csname LTb\endcsname%
      \put(4114,1537){\makebox(0,0)[r]{\strut{}ENAAM ($\gamma^{\rm max} = $ 10 MB)}}%
      \csname LTb\endcsname%
      \put(4114,1317){\makebox(0,0)[r]{\strut{}ENAAM ($\gamma^{\rm max} = $ 5 MB)}}%
      \csname LTb\endcsname%
      \put(4114,1097){\makebox(0,0)[r]{\strut{}DETA-R ($\gamma^{\rm max} = $ 10 MB)}}%
      \csname LTb\endcsname%
      \put(4114,877){\makebox(0,0)[r]{\strut{}DETA-R ($\gamma^{\rm max} = $ 5 MB)}}%
    }%
    \gplbacktext
    \put(0,0){\includegraphics{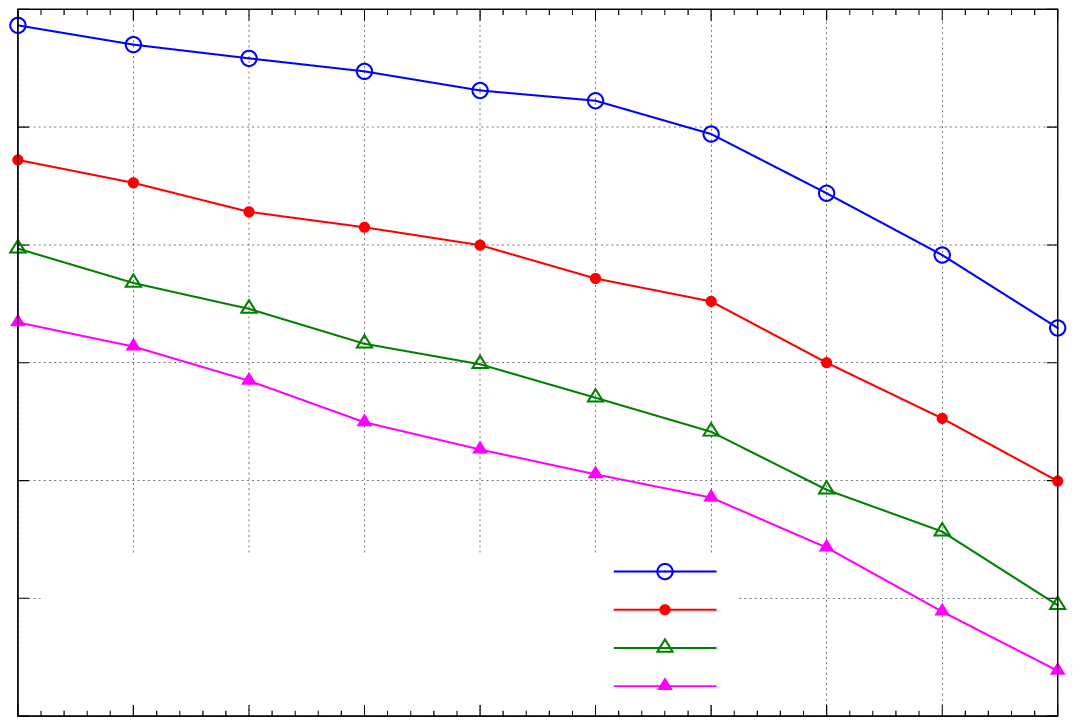}}%
    \gplfronttext
  \end{picture}%
\endgroup

%% file: clusterplot_multibs.tex
\begingroup
  \makeatletter
  \providecommand\color[2][]{%
    \GenericError{(gnuplot) \space\space\space\@spaces}{%
      Package color not loaded in conjunction with
      terminal option `colourtext'%
    }{See the gnuplot documentation for explanation.%
    }{Either use 'blacktext' in gnuplot or load the package
      color.sty in LaTeX.}%
    \renewcommand\color[2][]{}%
  }%
  \providecommand\includegraphics[2][]{%
    \GenericError{(gnuplot) \space\space\space\@spaces}{%
      Package graphicx or graphics not loaded%
    }{See the gnuplot documentation for explanation.%
    }{The gnuplot epslatex terminal needs graphicx.sty or graphics.sty.}%
    \renewcommand\includegraphics[2][]{}%
  }%
  \providecommand\rotatebox[2]{#2}%
  \@ifundefined{ifGPcolor}{%
    \newif\ifGPcolor
    \GPcolortrue
  }{}%
  \@ifundefined{ifGPblacktext}{%
    \newif\ifGPblacktext
    \GPblacktexttrue
  }{}%
  \let\gplgaddtomacro\g@addto@macro
  \gdef\gplbacktext{}%
  \gdef\gplfronttext{}%
  \makeatother
  \ifGPblacktext
    \def\colorrgb#1{}%
    \def\colorgray#1{}%
  \else
    \ifGPcolor
      \def\colorrgb#1{\color[rgb]{#1}}%
      \def\colorgray#1{\color[gray]{#1}}%
      \expandafter\def\csname LTw\endcsname{\color{white}}%
      \expandafter\def\csname LTb\endcsname{\color{black}}%
      \expandafter\def\csname LTa\endcsname{\color{black}}%
      \expandafter\def\csname LT0\endcsname{\color[rgb]{1,0,0}}%
      \expandafter\def\csname LT1\endcsname{\color[rgb]{0,1,0}}%
      \expandafter\def\csname LT2\endcsname{\color[rgb]{0,0,1}}%
      \expandafter\def\csname LT3\endcsname{\color[rgb]{1,0,1}}%
      \expandafter\def\csname LT4\endcsname{\color[rgb]{0,1,1}}%
      \expandafter\def\csname LT5\endcsname{\color[rgb]{1,1,0}}%
      \expandafter\def\csname LT6\endcsname{\color[rgb]{0,0,0}}%
      \expandafter\def\csname LT7\endcsname{\color[rgb]{1,0.3,0}}%
      \expandafter\def\csname LT8\endcsname{\color[rgb]{0.5,0.5,0.5}}%
    \else
      \def\colorrgb#1{\color{black}}%
      \def\colorgray#1{\color[gray]{#1}}%
      \expandafter\def\csname LTw\endcsname{\color{white}}%
      \expandafter\def\csname LTb\endcsname{\color{black}}%
      \expandafter\def\csname LTa\endcsname{\color{black}}%
      \expandafter\def\csname LT0\endcsname{\color{black}}%
      \expandafter\def\csname LT1\endcsname{\color{black}}%
      \expandafter\def\csname LT2\endcsname{\color{black}}%
      \expandafter\def\csname LT3\endcsname{\color{black}}%
      \expandafter\def\csname LT4\endcsname{\color{black}}%
      \expandafter\def\csname LT5\endcsname{\color{black}}%
      \expandafter\def\csname LT6\endcsname{\color{black}}%
      \expandafter\def\csname LT7\endcsname{\color{black}}%
      \expandafter\def\csname LT8\endcsname{\color{black}}%
    \fi
  \fi
    \setlength{\unitlength}{0.0500bp}%
    \ifx\gptboxheight\undefined%
      \newlength{\gptboxheight}%
      \newlength{\gptboxwidth}%
      \newsavebox{\gptboxtext}%
    \fi%
    \setlength{\fboxrule}{0.5pt}%
    \setlength{\fboxsep}{1pt}%
\begin{picture}(7200.00,5040.00)%
    \gplgaddtomacro\gplbacktext{%
      \csname LTb\endcsname%
      \put(682,704){\makebox(0,0)[r]{\strut{}$40$}}%
      \csname LTb\endcsname%
      \put(682,1518){\makebox(0,0)[r]{\strut{}$50$}}%
      \csname LTb\endcsname%
      \put(682,2332){\makebox(0,0)[r]{\strut{}$60$}}%
      \csname LTb\endcsname%
      \put(682,3147){\makebox(0,0)[r]{\strut{}$70$}}%
      \csname LTb\endcsname%
      \put(682,3961){\makebox(0,0)[r]{\strut{}$80$}}%
      \csname LTb\endcsname%
      \put(682,4775){\makebox(0,0)[r]{\strut{}$90$}}%
      \csname LTb\endcsname%
      \put(814,484){\makebox(0,0){\strut{}$1$}}%
      \csname LTb\endcsname%
      \put(1479,484){\makebox(0,0){\strut{}$2$}}%
      \csname LTb\endcsname%
      \put(2145,484){\makebox(0,0){\strut{}$3$}}%
      \csname LTb\endcsname%
      \put(2810,484){\makebox(0,0){\strut{}$4$}}%
      \csname LTb\endcsname%
      \put(3476,484){\makebox(0,0){\strut{}$5$}}%
      \csname LTb\endcsname%
      \put(4141,484){\makebox(0,0){\strut{}$6$}}%
      \csname LTb\endcsname%
      \put(4807,484){\makebox(0,0){\strut{}$7$}}%
      \csname LTb\endcsname%
      \put(5472,484){\makebox(0,0){\strut{}$8$}}%
      \csname LTb\endcsname%
      \put(6138,484){\makebox(0,0){\strut{}$9$}}%
      \csname LTb\endcsname%
      \put(6803,484){\makebox(0,0){\strut{}$10$}}%
    }%
    \gplgaddtomacro\gplfronttext{%
      \csname LTb\endcsname%
      \put(176,2739){\rotatebox{-270}{\makebox(0,0){\strut{}Mean energy savings $[\%]$}}}%
      \put(3808,154){\makebox(0,0){\strut{}Cluster size, $|O_i|$}}%
      \csname LTb\endcsname%
      \put(5038,4602){\makebox(0,0)[r]{\strut{}ENAAM ($\gamma^{\rm max} = $ 10 MB, $\eta = 0 $)}}%
      \csname LTb\endcsname%
      \put(5038,4382){\makebox(0,0)[r]{\strut{}ENAAM ($\gamma^{\rm max} = $ 5 MB, $\eta = 0 $)}}%
      \csname LTb\endcsname%
      \put(5038,4162){\makebox(0,0)[r]{\strut{}DETA-R ($\gamma^{\rm max} =$ 10 MB, $\eta = 0 $)}}%
      \csname LTb\endcsname%
      \put(5038,3942){\makebox(0,0)[r]{\strut{}DETA-R ($\gamma^{\rm max} =$ 5 MB, $\eta = 0 $)}}%
    }%
    \gplbacktext
    \put(0,0){\includegraphics{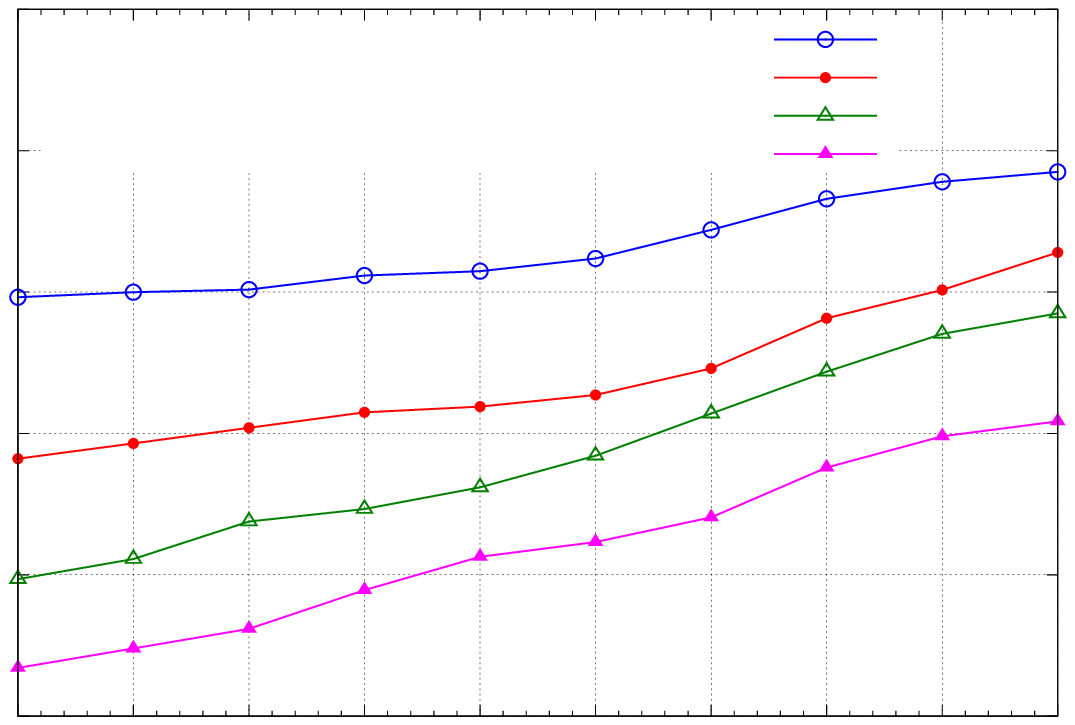}}%
    \gplfronttext
  \end{picture}%
\endgroup

%% file: weight_plot.tex
\begingroup
  \makeatletter
  \providecommand\color[2][]{%
    \GenericError{(gnuplot) \space\space\space\@spaces}{%
      Package color not loaded in conjunction with
      terminal option `colourtext'%
    }{See the gnuplot documentation for explanation.%
    }{Either use 'blacktext' in gnuplot or load the package
      color.sty in LaTeX.}%
    \renewcommand\color[2][]{}%
  }%
  \providecommand\includegraphics[2][]{%
    \GenericError{(gnuplot) \space\space\space\@spaces}{%
      Package graphicx or graphics not loaded%
    }{See the gnuplot documentation for explanation.%
    }{The gnuplot epslatex terminal needs graphicx.sty or graphics.sty.}%
    \renewcommand\includegraphics[2][]{}%
  }%
  \providecommand\rotatebox[2]{#2}%
  \@ifundefined{ifGPcolor}{%
    \newif\ifGPcolor
    \GPcolortrue
  }{}%
  \@ifundefined{ifGPblacktext}{%
    \newif\ifGPblacktext
    \GPblacktexttrue
  }{}%
  \let\gplgaddtomacro\g@addto@macro
  \gdef\gplbacktext{}%
  \gdef\gplfronttext{}%
  \makeatother
  \ifGPblacktext
    \def\colorrgb#1{}%
    \def\colorgray#1{}%
  \else
    \ifGPcolor
      \def\colorrgb#1{\color[rgb]{#1}}%
      \def\colorgray#1{\color[gray]{#1}}%
      \expandafter\def\csname LTw\endcsname{\color{white}}%
      \expandafter\def\csname LTb\endcsname{\color{black}}%
      \expandafter\def\csname LTa\endcsname{\color{black}}%
      \expandafter\def\csname LT0\endcsname{\color[rgb]{1,0,0}}%
      \expandafter\def\csname LT1\endcsname{\color[rgb]{0,1,0}}%
      \expandafter\def\csname LT2\endcsname{\color[rgb]{0,0,1}}%
      \expandafter\def\csname LT3\endcsname{\color[rgb]{1,0,1}}%
      \expandafter\def\csname LT4\endcsname{\color[rgb]{0,1,1}}%
      \expandafter\def\csname LT5\endcsname{\color[rgb]{1,1,0}}%
      \expandafter\def\csname LT6\endcsname{\color[rgb]{0,0,0}}%
      \expandafter\def\csname LT7\endcsname{\color[rgb]{1,0.3,0}}%
      \expandafter\def\csname LT8\endcsname{\color[rgb]{0.5,0.5,0.5}}%
    \else
      \def\colorrgb#1{\color{black}}%
      \def\colorgray#1{\color[gray]{#1}}%
      \expandafter\def\csname LTw\endcsname{\color{white}}%
      \expandafter\def\csname LTb\endcsname{\color{black}}%
      \expandafter\def\csname LTa\endcsname{\color{black}}%
      \expandafter\def\csname LT0\endcsname{\color{black}}%
      \expandafter\def\csname LT1\endcsname{\color{black}}%
      \expandafter\def\csname LT2\endcsname{\color{black}}%
      \expandafter\def\csname LT3\endcsname{\color{black}}%
      \expandafter\def\csname LT4\endcsname{\color{black}}%
      \expandafter\def\csname LT5\endcsname{\color{black}}%
      \expandafter\def\csname LT6\endcsname{\color{black}}%
      \expandafter\def\csname LT7\endcsname{\color{black}}%
      \expandafter\def\csname LT8\endcsname{\color{black}}%
    \fi
  \fi
    \setlength{\unitlength}{0.0500bp}%
    \ifx\gptboxheight\undefined%
      \newlength{\gptboxheight}%
      \newlength{\gptboxwidth}%
      \newsavebox{\gptboxtext}%
    \fi%
    \setlength{\fboxrule}{0.5pt}%
    \setlength{\fboxsep}{1pt}%
\begin{picture}(7200.00,5040.00)%
    \gplgaddtomacro\gplbacktext{%
      \csname LTb\endcsname%
      \put(682,849){\makebox(0,0)[r]{\strut{}$20$}}%
      \csname LTb\endcsname%
      \put(682,1576){\makebox(0,0)[r]{\strut{}$30$}}%
      \csname LTb\endcsname%
      \put(682,2303){\makebox(0,0)[r]{\strut{}$40$}}%
      \csname LTb\endcsname%
      \put(682,3030){\makebox(0,0)[r]{\strut{}$50$}}%
      \csname LTb\endcsname%
      \put(682,3757){\makebox(0,0)[r]{\strut{}$60$}}%
      \csname LTb\endcsname%
      \put(682,4484){\makebox(0,0)[r]{\strut{}$70$}}%
      \csname LTb\endcsname%
      \put(814,484){\makebox(0,0){\strut{}$0$}}%
      \csname LTb\endcsname%
      \put(1479,484){\makebox(0,0){\strut{}$0.1$}}%
      \csname LTb\endcsname%
      \put(2145,484){\makebox(0,0){\strut{}$0.2$}}%
      \csname LTb\endcsname%
      \put(2810,484){\makebox(0,0){\strut{}$0.3$}}%
      \csname LTb\endcsname%
      \put(3476,484){\makebox(0,0){\strut{}$0.4$}}%
      \csname LTb\endcsname%
      \put(4141,484){\makebox(0,0){\strut{}$0.5$}}%
      \csname LTb\endcsname%
      \put(4807,484){\makebox(0,0){\strut{}$0.6$}}%
      \csname LTb\endcsname%
      \put(5472,484){\makebox(0,0){\strut{}$0.7$}}%
      \csname LTb\endcsname%
      \put(6138,484){\makebox(0,0){\strut{}$0.8$}}%
      \csname LTb\endcsname%
      \put(6803,484){\makebox(0,0){\strut{}$0.9$}}%
    }%
    \gplgaddtomacro\gplfronttext{%
      \csname LTb\endcsname%
      \put(176,2739){\rotatebox{-270}{\makebox(0,0){\strut{}Mean energy savings $[\%]$}}}%
      \put(3808,154){\makebox(0,0){\strut{}Weight, $\eta$}}%
      \csname LTb\endcsname%
      \put(4114,1537){\makebox(0,0)[r]{\strut{}ENAAM ($\gamma^{\rm max} = $ 10 MB)}}%
      \csname LTb\endcsname%
      \put(4114,1317){\makebox(0,0)[r]{\strut{}ENAAM ($\gamma^{\rm max} = $ 5 MB)}}%
      \csname LTb\endcsname%
      \put(4114,1097){\makebox(0,0)[r]{\strut{}DETA-R ($\gamma^{\rm max} = $ 10 MB)}}%
      \csname LTb\endcsname%
      \put(4114,877){\makebox(0,0)[r]{\strut{}DETA-R ($\gamma^{\rm max} = $ 5 MB)}}%
    }%
    \gplbacktext
    \put(0,0){\includegraphics{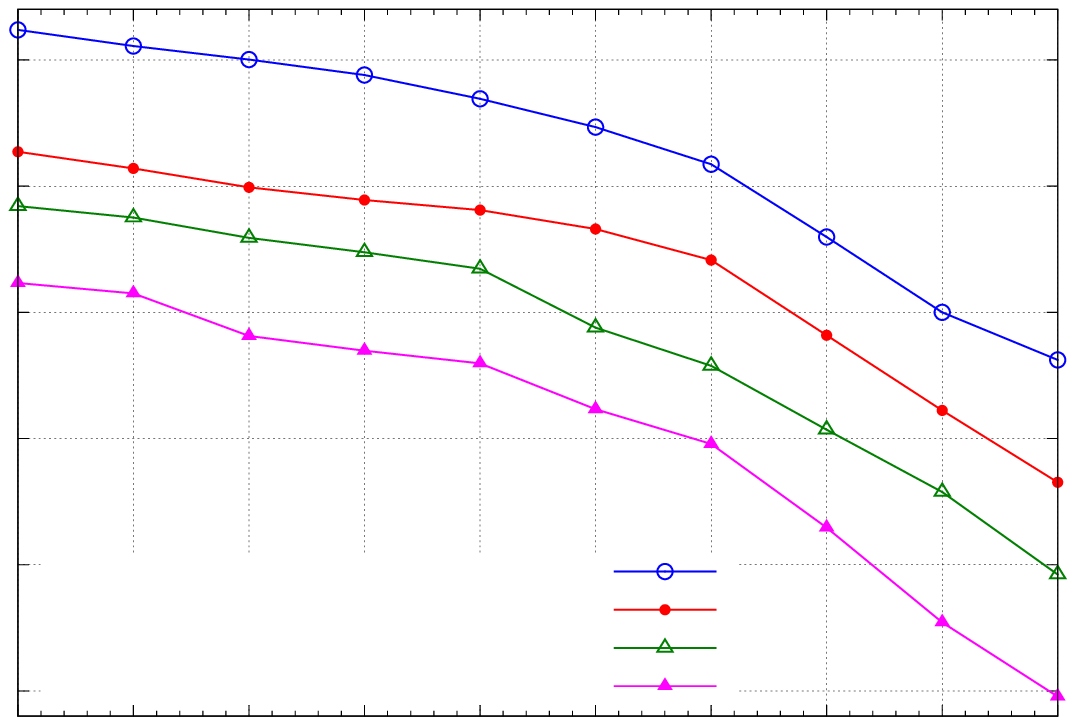}}%
    \gplfronttext
  \end{picture}%
\endgroup